\documentclass[aps,prd,amsmath,10pt,amssymb,twocolumn,showpacs,superscriptaddress,raggedbottom,preprintnumbers]{revtex4-1}

\usepackage{graphicx}
\usepackage{dcolumn}
\usepackage{bm}
\usepackage{hyperref}
\usepackage[utf8]{inputenc}
\usepackage[dvipsnames]{xcolor}

\graphicspath{{figures/}}

\begin{document}

\title{Simulating hadron test beams in liquid argon}

\author{Alexander Friedland}
 \email{alexfr@slac.stanford.edu}
 
\author{Shirley Weishi Li}%
 \email{shirleyl@slac.stanford.edu}
\affiliation{SLAC National Accelerator Laboratory, 2575 Sand Hill Road, Menlo Park, CA, 94025}

\date{October 12, 2020}

\begin{abstract}
Thorough modeling of the physics involved in liquid argon calorimetry is essential for accurately predicting the performance of DUNE and optimizing its design and analysis pipeline. At the fundamental level, it is essential to quantify the detector response to individual hadrons---protons, charged pions, and neutrons---at different injection energies. We report such a simulation, analyzed under different assumptions about event reconstruction, such as particle identification and neutron detection. The role of event containment is also quantified. The results of this simulation can help inform the ProtoDUNE test-beam data analysis, while also providing a framework for assessing the impact of various cross section uncertainties. 
\end{abstract}

\preprint{SLAC-PUB-17549}

\maketitle


\section{Motivations}
\label{sec:level1}

Energy resolution and the accuracy of energy scale calibration are essential characteristics for a neutrino detector operating in a broad-spectrum neutrino beam. Modeling these characteristics for the Deep Underground Neutrino Experiment (DUNE) is a nontrivial task. At the root of the problem is the nature of the final states produced when neutrinos of several-GeV energies interact with argon nuclei. These interactions can produce multiple hadrons of different types, which can, in turn, undergo subsequent interactions in the detector medium, distributing energy among even more particles. At first sight, by collecting all ionization charges, one should be able to measure all this energy calorimetrically. In reality, however, different particles create different amounts of detectable charge per unit energy lost, and some energy goes into invisible channels, such as nuclear breakup. As an extreme case, some or all neutrons may be altogether missed. Thus, having an accurate model for the detector response to each particle type is essential for optimal detector performance. 

Given the complexity of the problem, a consistent way to study it is to simulate a large number of fully developed neutrino events~\cite{Abi:2020evt,Friedland:2018vry}. The simulation pipeline in this approach combines a code modeling the primary neutrino interaction with another one propagating all resulting particles through the liquid argon medium. For the first code, one can use {\tt GENIE}, {\tt GiBUU}, or another event generator. For the second, the choices are {\tt GEANT4} or {\tt FLUKA}, both of which model not only ionization losses, but also any subsequent hadronic and electromagnetic interactions of all particles in the detector. The process needs to be repeated for different flavors of the incoming neutrino and a range of energy of interest. The result is a set of migration matrices describing probabilities connecting true and reconstructed energies. These matrices are an essential input for any analysis of oscillation sensitivity.

All this computer-intensive process is necessary just to characterize energy resolution in the case of baseline assumptions about the detector performance. If one wishes to investigate the impact of various changes to the reconstruction procedure, one needs to rerun the entire simulation pipeline. For example, one may wish to vary detection thresholds, exclude certain particle types, or investigate the impact of various particle identification (PID) assumptions. In each case, one obtains a new set of migration matrices, which then can be used for oscillation studies. An example study following this approach is presented in Ref.~\cite{Friedland:2018vry}, where we considered several model assumptions about the detector performance, specifically, on the values of particle detection thresholds and the availability of accurate PID information. 

To gain more insight into the physics dictating neutrino detection in liquid argon, in this paper, we inject in our simulation volume individual hadron particles---protons, charged pions, neutrons---and investigate the detector response in each case. This should allow one to understand the role of each particle type in neutrino energy reconstruction. Of particular interest is to quantify the importance of reconstructing secondary neutrons.

There are several additional reasons to consider this study. First, our simulations yield ``virtual test-beam data'', which can be used to compare with the actual ProtoDUNE~\cite{Abi:2020mwi} test-beam data, an essential step to validating the entire simulation framework for the full events.

Second, it may also be used to devise sanity checks for the full event simulation results. Such checks are always necessary when one deals with large simulation frameworks with complex codes. 

Third, one can use the results on individual particles to create simplified, flexible codes, in which prescriptions describing the detection process are applied to the outputs of the neutrino event generator. This is the general philosophy of the {\tt FastMC} code employed in the DUNE CDR documents~\cite{Acciarri:2015uup,Acciarri:2016ooe}. We consider this approach to be very useful for certain problem types and far from being completely superseded by the full simulations. In connection with this point, it is extremely important to establish under which conditions the reconstructed energy for a given hadron type may be described by a Gaussian.

In our study here, we do not directly focus on detector signatures of electrons, muons, or gamma rays. The reason is based on our findings in Ref.~\cite{Friedland:2018vry}. Despite the different event topologies---muons leave long tracks, while electrons and gamma rays create electromagnetic showers---in all three cases, the total ionization charge was found to be in close correspondence with the true particle energy. Thus, the experimental energy resolution for these particles will likely be controlled by the reconstruction algorithm performance and not by physical processes in particle propagation, which are the focus of the present study.

The presentation is organized as follows. Section~\ref{sect:overview} presents an overview of our simulation framework and a list of specific reconstruction assumptions considered in the paper. Section~\ref{sect:results} presents the simulation results for each hadron type. Section~\ref{sect:protons} treats protons and also explains the reconstruction procedure. Section~\ref{sect:pions} treats charged pions while Sec.~\ref{sect:neutrons} is devoted to the study of neutrons. Section~\ref{sect:limitedvolume} explores the impact of limited detector volume. Section~\ref{sect:discussion} discusses some consequences of the results of our study, including the physics dictating the energy resolution and possible applications to the development of simplified codes. Finally, Sec.~\ref{sec:conclusions} summarizes our main findings.


\section{Simulation overview}
\label{sect:overview}

Following Ref.~\cite{Friedland:2018vry}, our simulations here also employ {\tt FLUKA}~\cite{Bohlen:2014buj, Ferrari:2005zk} to model event development in liquid argon. {\tt FLUKA}---here we use version 2011.2x.6---is a publicly available, well-tested package that incorporates all relevant physics processes, such as ionization and radiative energy losses, hadronic inelastic interaction, and particle decays. Among its many strengths is a good description of MeV hadronic physics, as recently demonstrated by the ArgoNeuT experiment~\cite{Acciarri:2018myr}. 

As in Ref.~\cite{Friedland:2018vry}, we fully propagate all particles, including those produced in secondary interactions, but do not consider detector-specific effects, such as the finite lifetime of drifted charges, space charge distribution, wire spacing, electronic noise, or cosmogenic and radiogenic backgrounds. Such studies are beyond the scope of the present paper and will depend on specific detector configurations and performance characteristics. We are encouraged, however, by the extremely low levels of electronic noise in the ProtoDUNE-SP data~\cite{Abi:2020mwi} and assume that the reported issues with the space charge distribution will be adequately resolved.

\begin{figure}
    \centering
    \includegraphics[width=0.95\columnwidth]{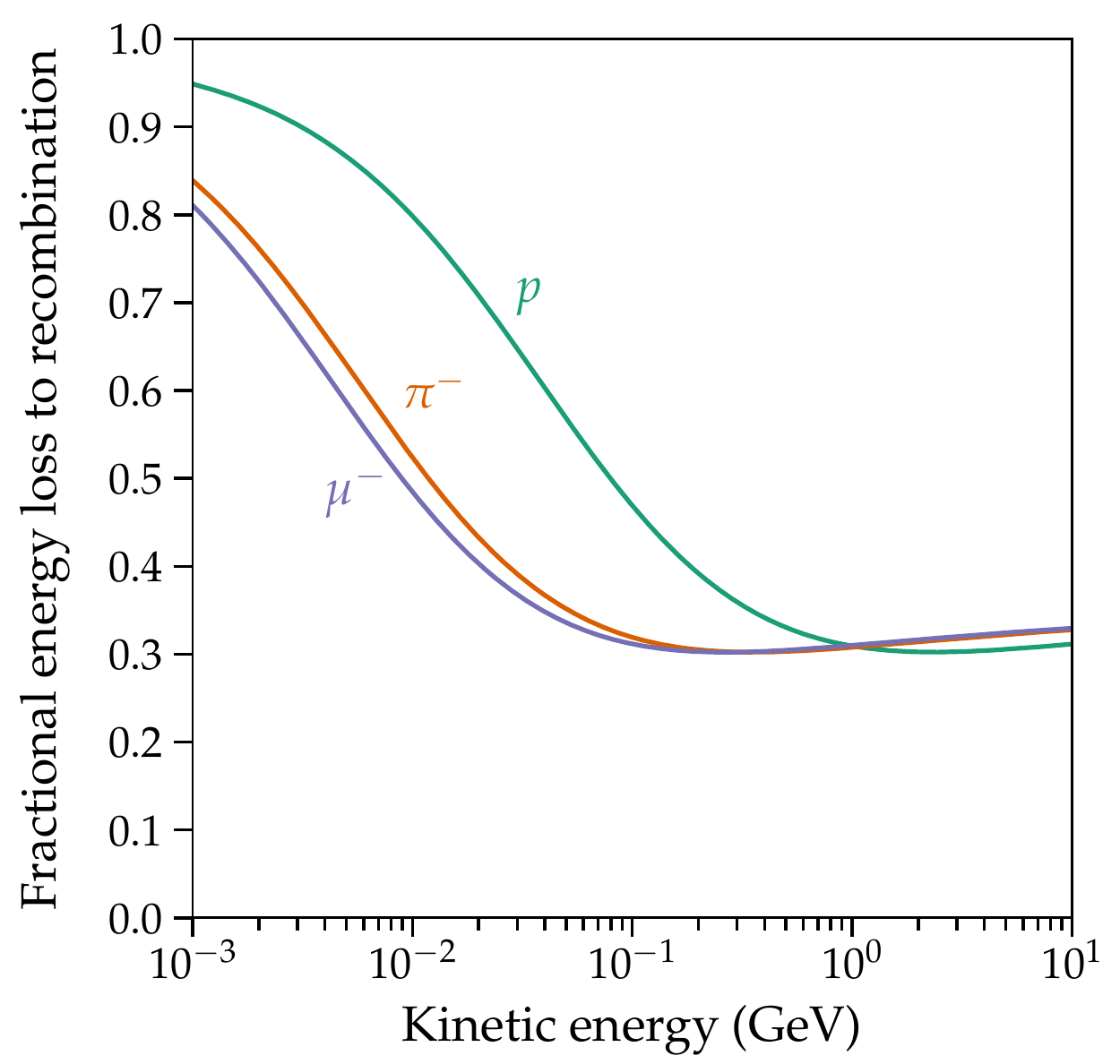}
    \caption{Fraction of the ionization charge lost to recombination as a function of the kinetic energy of the particle, for protons, charged pions, and muons.}
    \label{fig:recombination}
\end{figure}

Our emphasis at present is \emph{on assessing the physical impact of different reconstruction assumptions}. Specifically, we aim to elucidate the impact of good PID and neutron detection. We argued in Ref.~\cite{Friedland:2018vry} that these are crucial factors determining the accuracy of neutrino energy measurements in liquid argon. Here, we deconstruct the argument by considering the reconstruction process for each hadron type. Accordingly, we analyze three model scenarios:
\begin{enumerate}
     \item \emph{Best reconstruction}. One has PID information on all charged particles in an event and applies it to get the ionization energy loss along each trajectory. The detection thresholds are considered to be very low, motivated by the ArgoNeuT experiment.
     \item \emph{Charge-only reconstruction}. No PID information is available for any secondary particles in an event. One collects the total ionization charge and uses it to infer, statistically, the energy of the injected particle. 
     \item \emph{Charge-only, no neutrons}. In addition to the assumption of no PID, any energy imparted to neutrons at any stage in the process development is considered to be completely lost.
\end{enumerate}

\begin{figure*}
    \centering
    \includegraphics[width=0.95\columnwidth]{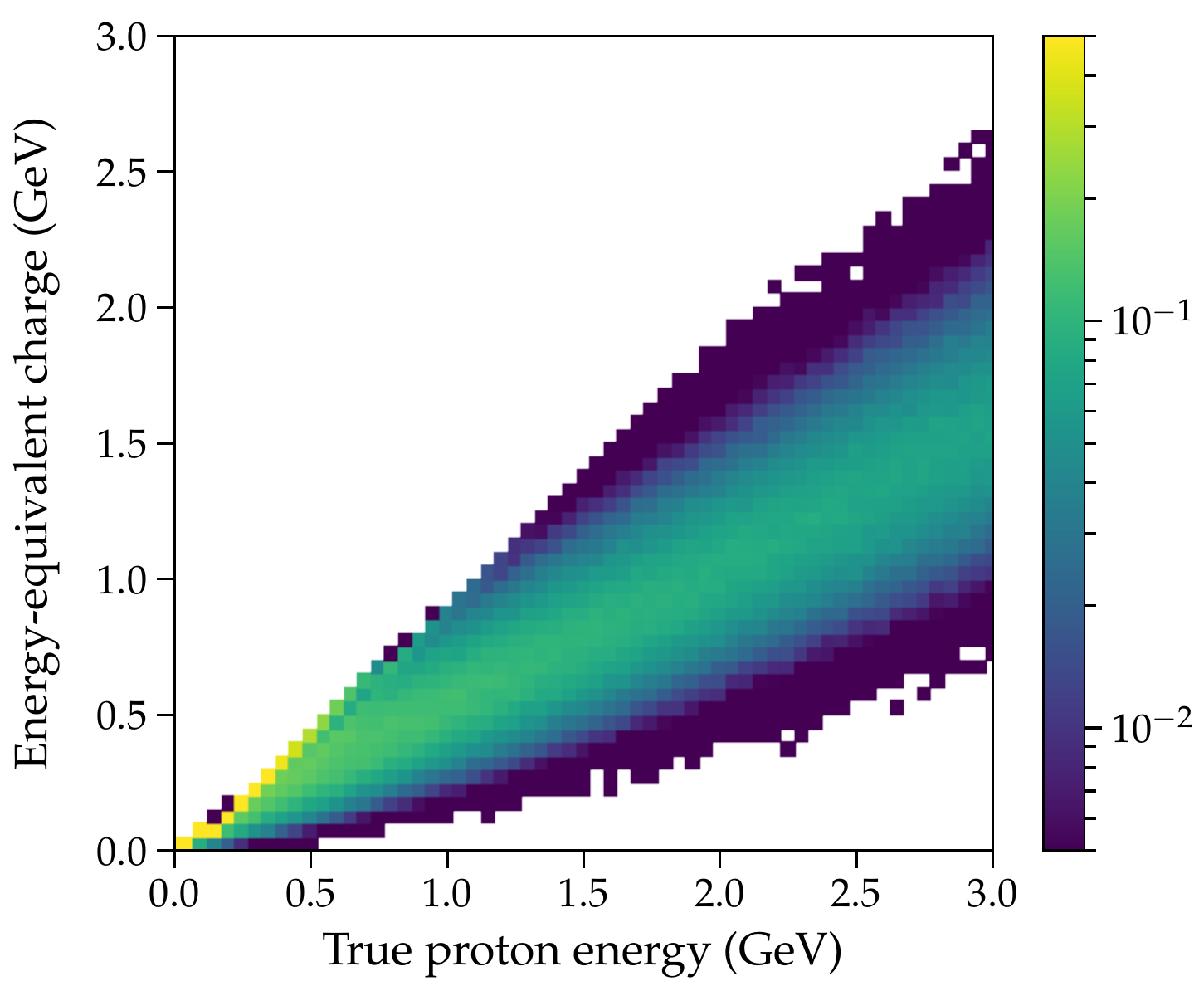}
    \includegraphics[width=0.95\columnwidth]{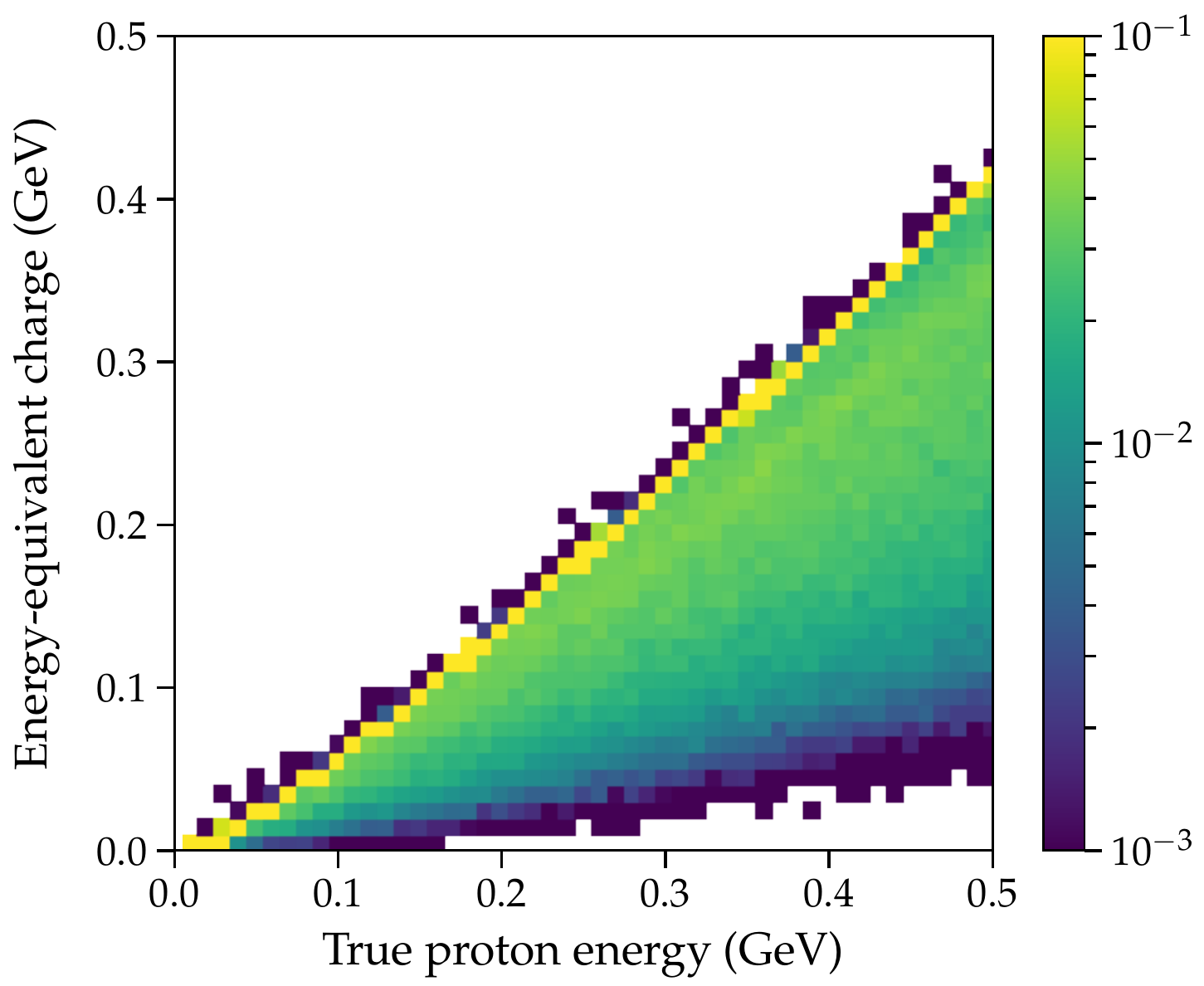}
    \caption{Graphical representation of the matrix connecting ionization charge $Q$ and the true energy of the injected proton $E_{tr}$, according to our simulations. \emph{Left}: $E_{tr}$ from 0 to 3 GeV, \emph{right}: $E_{tr}$ from 0 to 0.5 GeV at finer numerical sampling. The color of each square indicates the probability of obtaining the corresponding interval of charge $Q$ given the value of $E_{tr}$.}
    \label{fig:QvsEtrue}
\end{figure*}

The first two scenarios were already considered in Ref.~\cite{Friedland:2018vry}. The second method is described in detail in Refs.~\cite{Sorel:2014rka,DeRomeri:2016qwo} and is currently accepted within the DUNE Collaboration as a way of treating the hadronic system~\cite{Grant2018DPFtalk}. The two scenarios represent extreme approaches to the treatment of PID information and thus allow us to quantify the impact of good PID on energy resolution. Realistic reconstruction efforts should produce an answer that is intermediate between the two scenarios, by reconstructing some of the particles, and the goal is to inform such efforts on how much can be gained by improved reconstruction.

The significance of these assumptions becomes obvious when one considers the anatomy of a hadronic event in liquid argon. Slower particles create more dense charge tracks, which, in turn, leads to more charge loss to recombination. Thus, a relationship between the detected charge, $dQ/dx$, and the true ionization energy loss, $dE/dx$, depends on the particle type.  This is illustrated in Fig.~\ref{fig:recombination}, which shows what fraction of the ionization charge, created by protons, muons, and charged pions, of varying energies, is lost to recombination. The main physical effect here is that protons, being more massive, deposit more dense charge tracks, where more charge recombination takes place. As a corollary, protons yield less observable charge per unit energy lost than charged pions or muons of the same kinetic energy.

The third scenario is also introduced here as a deliberately extreme case. It allows us to quantify the importance of detecting neutrons. Neutrons, being electrically neutral, do not leave tracks at all. Their presence can be detected by the secondary ionizing particles they produce in their interactions. Since these interactions occur some distance away from their starting points, one ends up with secondary proton and pion tracks separated from the main event, and with a spray of small charge deposits created by the de-excitation gamma rays, from multiple disrupted nuclei, undergoing repeated Compton scattering~\cite{Friedland:2018vry,Castiglioni:2020tsu}. The detectability of these different signatures will depend on a number of experimental factors---including reconstruction performance, containment issues due to finite detector volume, electronic noise, radiogenic and cosmogenic backgrounds---and may be different for the near and far detectors. Nevertheless, it is highly reassuring that ArgoNeuT has demonstrated good ability to detect charge ``blips" from nuclear de-excitations~\cite{Acciarri:2018myr}, while MicroBooNE has done the same for $^{39}$Ar beta decay products~\cite{MicroBooNE:2018jag}. So, once again, a realistic treatment of neutrons is expected to be somewhere between scenarios 1 and 3.

In summary, the three scenarios outlined above are chosen to quantify the impact of the two main physical aspects of event reconstruction: the efficiencies fo PID and of neutron detection.


\section{Simulation results}
\label{sect:results}


\subsection{Protons}
\label{sect:protons}

\begin{figure*}
	\begin{center}
        \includegraphics[width=\textwidth]{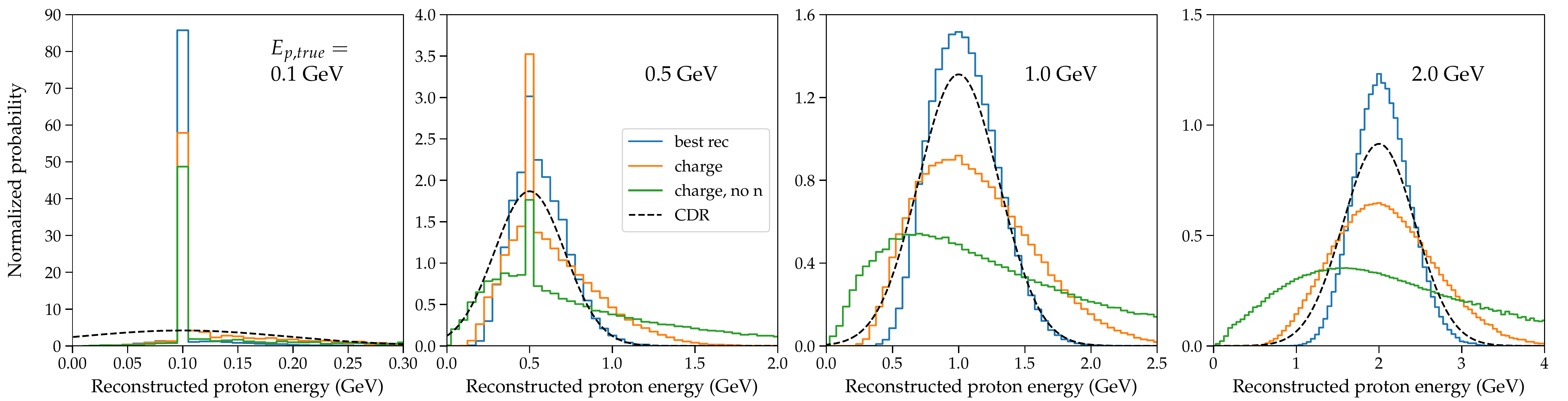}
        \caption{Distributions of proton reconstructed energies, for four representative values of the true energy, $E_p=0.1$, $0.5$, $1.0$, and $2.0$ GeV. Three different reconstruction scenarios are considered: (1) full PID information is available (\emph{blue}), (2) only total ionization charge (\emph{orange}), and (3) total ionization charge with neutrons undetected (\emph{green}). For comparison, the dashed curve shows the resolution assumed in the DUNE CDR document~\cite{Acciarri:2015uup,Acciarri:2016ooe}.}
        \label{fig:proton4panels}
    \end{center}
\end{figure*}

We begin by considering proton beams. We will use this case to describe our simulation procedure and to explain how we use its results to model energy resolution.

First, we generate our simulation dataset, which is used in each of our three scenarios. For this, we inject protons of energies from 0.01 to 3.0~GeV into unlimited liquid argon volume and model the full event development in each case. The effects of finite detector size deserve a dedicated discussion and we will address them in Sec.~\ref{sect:limitedvolume} below. Between 0.1 and 3.0 GeV, we sample proton energy values in 0.05 GeV intervals. To better characterize the resolution at low energies, we also generate the second dataset with 0.01 GeV energy spacing. For each value of the true energy, $E_{tr}$, we generate $10^4$ events.

For definiteness, let us for now specialize to the total-charge study (scenario 2). The simulation dataset tells us, for a given value of the true proton energy, $E_{tr}$, the probability $P(Q|E_{tr})$ of measuring charge $Q$. Discretizing (binning) the $Q$ values, we obtain a matrix of probabilities connecting $E_{tr}$ and $Q$.  Explicitly, the matrix element $P(Q^{(j)}|E_{tr}^{(i)})$ equals the number of events that landed in bin $Q^{(j)}$ divided by the number of simulations with $E_{tr}^{(i)}$. This matrix, in the graphical form, is presented in Fig.~\ref{fig:QvsEtrue}. The left panel covers the full range of $E_{tr}$, up to 3 GeV; the right panel depicts the finely sampled low-energy dataset, covering $E_{tr}$ values up to 0.5 GeV. The width of the bins in $Q$ is the same as the corresponding $E_{tr}$ bins: 0.05 GeV in the first simulation and 0.01 GeV in the second one.

Now, suppose we use this simulation dataset to analyze a new event, created by a proton with an unknown value of $E_{tr}$. Given the value of $Q$ measured for this event, we can use our matrix as a lookup table, to obtain the probability $P(E_{rec}|Q)$ that the event was created by a proton with energy $E_{rec}$. Explicitly, $P(E_{rec}^{(j)}|Q^{(i)})$ is equal to the number of times charge $Q^{(i)}$ was obtained in the simulation with proton energy $E_{rec}^{(j)}$ divided by the total number of times charge $Q^{(i)}$ was obtained for all energies in our simulation set. 

What we just described provides a reconstruction procedure, which allows one to go from the detected value of $Q$ to $E_{rec}$. The requirement of unbiased reconstruction is assured by construction, since our proton energy values are drawn from a flat distribution. This procedure is designed to capture the reconstruction problem for an actual (non-test-beam) experiment, where, in a most general case, one does not have any prior information on the distribution from which proton energies are drawn. For example, one may have a proton track in the middle of a complicated neutrino event. 

In summary, while reading the matrix in Fig.~\ref{fig:QvsEtrue} ``vertically” gives us a probability distribution  of ionization charge $Q$ for a proton of a known $E_{tr}$, reading the same matrix ``horizontally”, with appropriate normalization, provides an unbiased reconstruction procedure.

We now have all ingredients to describe our full energy resolution modeling pipeline. To this end, it is important to precisely define what we mean by \emph{energy resolution.} Loosely speaking, this term could simply refer to the width of the $Q$ distribution that is obtained when a beam of a given energy $E_{tr}$ is shot into the detection volume. More accurately, energy resolution in this paper will describe the width of the distribution $P(E_{rec}|E_{tr})$, of reconstructed energies $E_{rec}$ obtained starting with $E_{tr}$ and reconstructing each event independently, using the lookup procedure described above.

To find the probability distribution of reconstructed energies, $P(E_{rec}|E_{tr})$, we must sum (integrate) over all $Q$ values that can be obtained in the intermediate step:
\begin{eqnarray}
P(E_{rec}|E_{tr}) = \int dQ P(E_{rec}|Q) P(Q|E_{tr}) .
\end{eqnarray}{}

It can be straightforwardly shown that, if the charge distribution is Gaussian, 
\begin{equation}
    P(Q|E_{tr}) = \frac{1}{\sqrt{2\pi} \sigma} \exp\left[-\frac{(Q-E_{tr}f)^2}{2\sigma^2}\right],
\end{equation}
where $f$ is the fraction of energy that on average goes into charge, the resulting distribution of $E_{rec}$ is also Gaussian, with the width $\sqrt{2}\sigma/f$:
\begin{eqnarray}
    P(E_{rec}|E_{tr}) &=& \int dQ P(E_{rec}|Q) P(Q|E_{tr}) \nonumber\\
    &=& \frac{f}{2\sqrt{\pi} \sigma} \exp\left[-\frac{(E_{rec}-E_{tr})^2 f^2}{4\sigma^2}\right].
\end{eqnarray}
Here the probability distribution $P(E_{rec}|Q)$ is given by
\begin{equation}
    P(E_{rec}|Q) = \frac{f}{\sqrt{2\pi} \sigma} \exp\left[-\frac{(Q-E_{rec}f)^2}{2\sigma^2}\right],
\end{equation}
which is normalized to 1. In a general case, however, the distributions for $E_{rec}$ and $Q$ do not follow the same functional form.

The application of this procedure to the other two reconstruction methods is now straightforward. For the simulation with no neutrons, all charges created downstream of any neutron are discarded, with the rest of the procedure unaffected. In the best-reconstruction case, to each track in the event, we apply a charge recombination correction factor that is a function of its PID. The resulting distribution of the ``modified charged" is used in place of $Q$. 

\begin{figure*}
	\begin{center}
        \includegraphics[width=\textwidth]{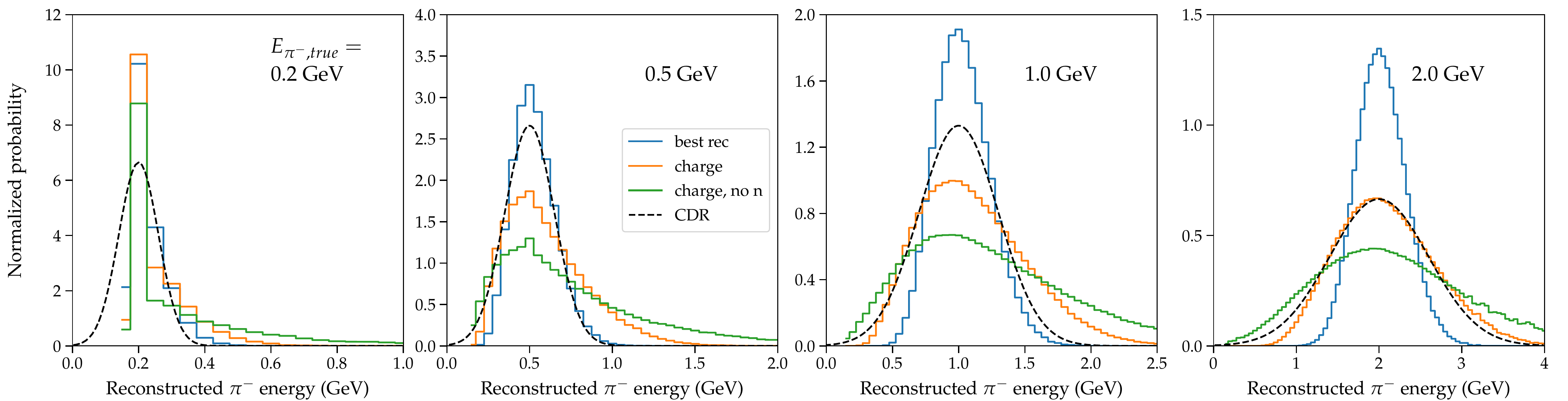}
        \caption{Same as Fig.~\ref{fig:proton4panels}, but for the $\pi^-$ reconstructed energies.}
        \label{fig:pionn4panels}
    \end{center}
\end{figure*} 

Figure~\ref{fig:proton4panels} shows the result of applying this procedure to our simulation set. Four representative values of the true proton energy are considered: 0.1, 0.5, 1.0, and 2.0 GeV. We see that the character of the distribution changes as one goes from low to high energy values: at 2 GeV, the $E_{rec}$ distributions are well described by Gaussians, while at 0.1 GeV the distribution is dominated by a sharp spike, where essentially all proton energy is recovered. The 0.5~GeV represents a transition between these regimes. This transition is also clearly seen in Fig.~\ref{fig:QvsEtrue}, where the yellow diagonal points represent unscattered protons.

The change from the unscattered to multiply-scattered regimes is dictated by the mean free path of hadronic collisions. It will prove crucial for our discussion in Sec.~\ref{sect:discussion} below. But first, we turn to the corresponding results for the other hadrons.
 
 
\subsection{Charged pions}
\label{sect:pions}

Understanding the propagation of charged pions is also of direct relevance to DUNE calorimetry. As illustrated in Ref.~\cite{Friedland:2018vry}, interactions of 4 GeV neutrinos can create hadronic showers with multiple pions, with energies in the hundreds of MeV range. Even 1--2 GeV pions are not uncommon in such events. Therefore, it is certainly worth considering charged pion test beams, and indeed ProtoDUNE has collected such data.

In Fig.~\ref{fig:pionn4panels}, we simulate charged pion beams, with energies 0.2, 0.5, 1 and 2 GeV. The histograms in the figure correspond to $\pi^-$; for positively charged pions, the results are very similar. The reconstruction assumptions and the analysis are the same as considered earlier for protons. We see that the basic results for pions and protons are qualitatively similar: the distributions of reconstructed energies are non-Gaussian at the lowest energies and become Gaussian at higher energies. One notable quantitative difference is that the Gaussianity sets in at a smaller energy for pions than for protons. This has a natural physical explanation in terms of the ionization rates in the two cases. With energy loss having a $\propto v^{-2}$ leading velocity dependence, slower particles lose energy faster per unit distance traveled. Since protons of a few hundred MeV are nonrelativistic, their ionization rates are higher than for pions of the same kinetic energy. Thus, protons are more likely than pions to come to rest before undergoing hadronic interactions, and it is repeated hadronic interactions that create Gaussian distribution of reconstructed energies.


\subsection{Neutrons}
\label{sect:neutrons}

We have seen that neutron detection has a dramatic impact on the accuracy of the calorimetric energy reconstruction by liquid argon detectors. Let us now take a deeper dive into the subject by analyzing neutron propagation and interactions. 

First of all, one should be more precise about what is meant by neutron detection. As already mentioned in Sec.~\ref{sect:overview}, a neutron traveling through the liquid argon medium does not, by itself, create an ionization track. Its energy is lost via interaction with multiple argon nuclei, and it is through the secondary particles created in these interactions that the presence of the neutron can be revealed. Importantly, the secondary charged particles carry only a fraction of the original neutron energy---some of the energy is lost to nuclear breakup. Hence, a direct calorimetric measurement of the neutron energy is not possible. One recovers only part of the energy and uses a simulation-based model to infer the likely energy range of the original neutron.

\begin{figure}
	\begin{center}
        \includegraphics[width=\columnwidth]{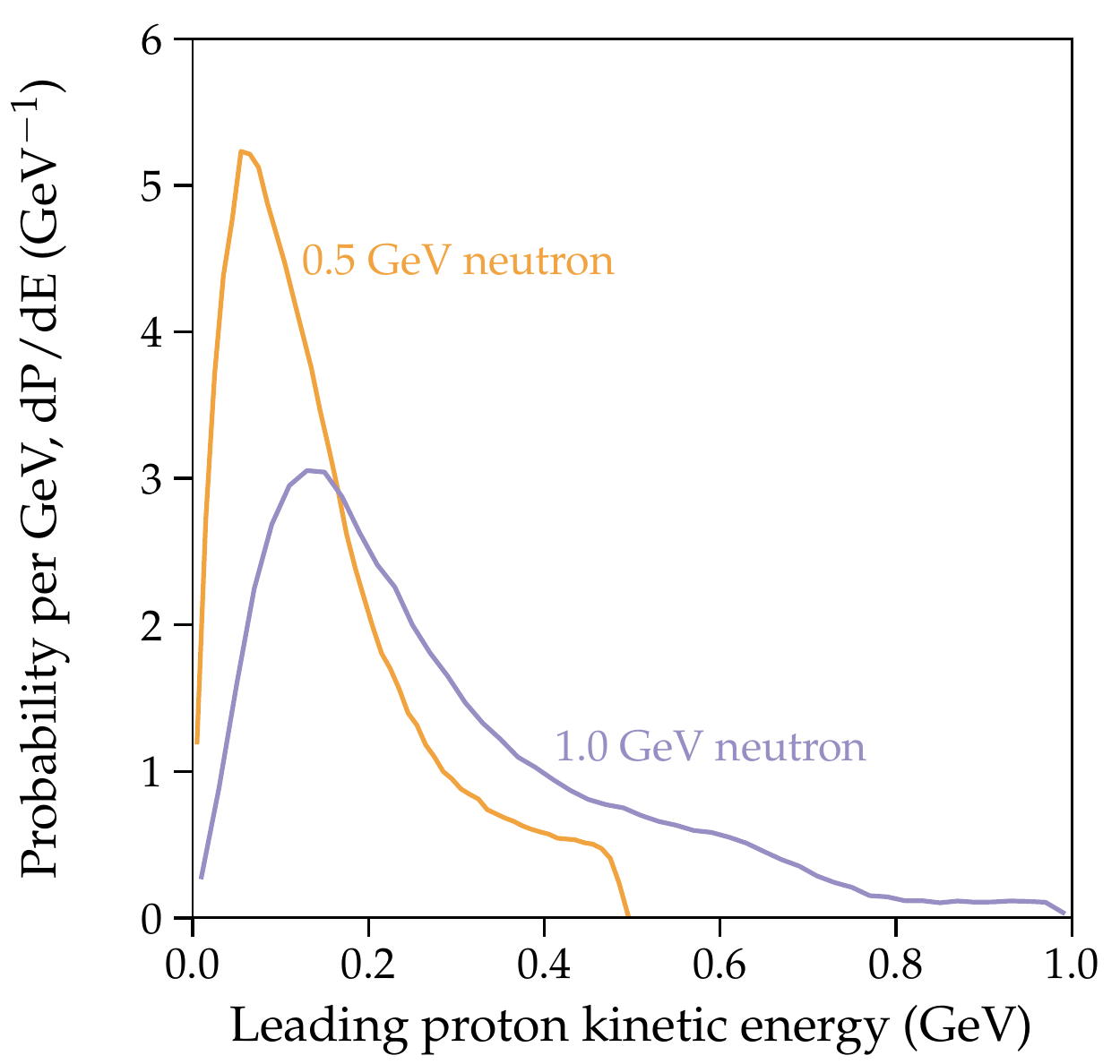}
        \caption{Kinetic energy distributions of the most energetic protons produced by 0.5 and 1.0~GeV neutrons.}
        \label{fig:nprodp}
    \end{center}
\end{figure}

\begin{figure*}
	\begin{center}
        \includegraphics[width=\textwidth]{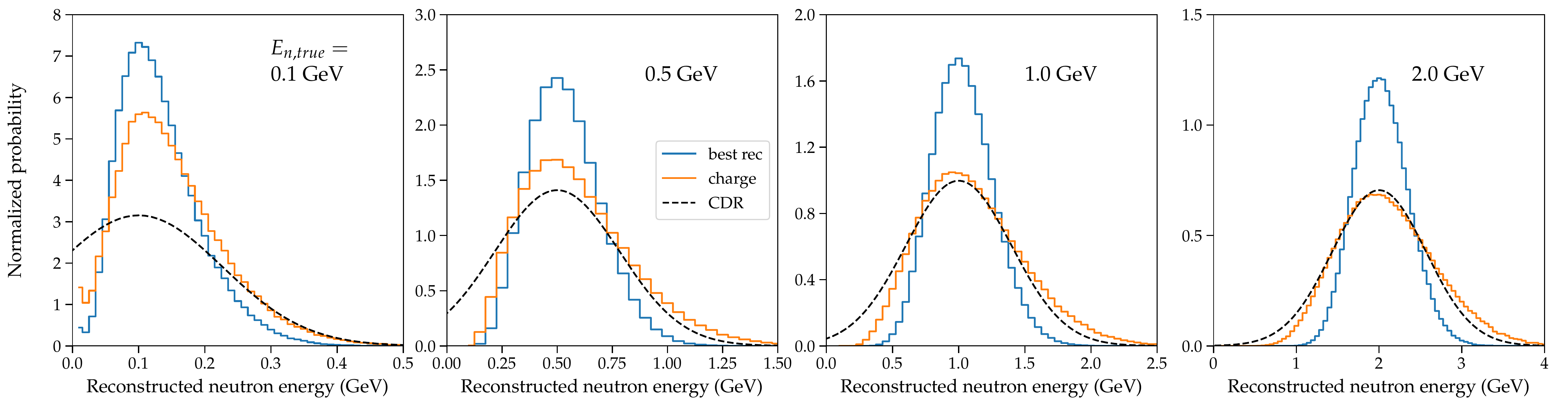}
        \caption{Distributions of the neutron reconstructed energies, for four representative values of the true energy, $E_n=0.1$, $0.5$, $1.0$, and $2.0$ GeV. Two different reconstruction scenarios are considered: (1) full PID information is available (\emph{blue}) and (2) only total ionization charge (\emph{orange}). For comparison, the dashed curve shows the resolution assumed in the DUNE CDR document.}
        \label{fig:neutron4panels}
    \end{center}
\end{figure*}

At a more detailed level, one has to consider the different signatures that can be created in neutron interactions. A neutron can excite an argon nucleus, or it can knock out one or more nucleons from it, leaving the daughter nucleus in an excited state. The de-excitation gammas undergo Compton scattering in the medium, and the recoil electrons leave small ionization charge deposits~\cite{Castiglioni:2020tsu}. Since a given neutron interacts with many argon nuclei in this way, many recoil electrons are scattered over an extended region. The resulting spray of such small charges, from many nuclear interactions, is, in principle, observable, as demonstrated by the ArgoNeuT analysis~\cite{Acciarri:2018myr}. 

A more prominent signature comes from energetic knockout products. In particular, a sufficiently energetic proton can create a distinct track that is detached from the main event. Such tracks can be identified as protons, thus enabling proper charge recombination correction. In Fig.~\ref{fig:nprodp}, we depict a spectrum of the leading (highest energy) protons created in propagation of neutrons of two starting kinetic energies: 0.5 and 1 GeV. Estimating the threshold for proton identification to be 30 MeV, we see that a large fraction of the knockout protons could be identified.

This remains true even at lower neutron energy. For example, for a 300 MeV neutron, on average, 34\% of the energy goes into knockout protons above the 30 MeV threshold, according to our simulations. An additional 4\% of the energy goes into protons below that threshold; 40\% is lost to nuclear breakup; 14\% goes into gammas; 4\% is imparted to heavy ions knocked out of the nuclei; 2\% goes to nuclear recoil; and, finally, 2\% goes to pions produced in hadronic collisions. Thus, the full energy budget is quite complicated, and the accuracy of energy reconstruction depends on how much of that energy can be recovered.

Three comments about these numbers are in order. First, the process is highly stochastic, and event-to-event variations are found to be large. For example, the energy fraction in the leading proton has a range of $38 \pm 24\%$. Second, the fractions obviously change with neutron energy. In particular, inelastic hadronic interactions become more prominent at higher energy. For 1-GeV neutrons, as much as 19\% of the energy goes into pions. Third, the components in subthreshold protons and in heavy ions require special attention.

The subthreshold protons are those for which the reconstruction algorithm is not able to identify a clear track. In such a case, they appear as part of the spray. Compared to Compton-recoil electrons, however, which make up a lot of the charge ``blips" in the spray, these low-energy hadrons are subject to larger charge recombination. Thus, if one wished to use the measured charge in the spray to improve the neutron energy reconstruction---compared to what is possible from the leading proton alone---the composition of the spray must be reliably understood. On the other hand, the charge blips due to low-energy protons should have a significantly higher charge concentration than the corresponding electron blips. The authors propose to investigate if reconstruction algorithms could be taught to distinguish proton and electron blips based on the charge concentration, even when a definite track could not be identified.

Similar considerations also apply to the heavy ion products. Our simulation shows that most of such ions are deuterons, with the average kinetic energy of 64 MeV at production. Taking the rates of charge recombination into account, a 64 MeV deuteron deposits as much charge as a 53 MeV proton, but with a $dE/dx$ energy loss rate that is approximately 50\% greater. Hence, it might in principle be possible to train the reconstruction software to also identify most deuterons, further improving energy reconstruction.

For all these reasons, detailed analysis of the nature of the small charge deposits represents a potentially promising way to improve energy resolution of the liquid argon technology. The necessary studies on subthreshold protons and deuteron ions could be carried out at MiniBooNE and ProtoDUNE.

Finally, all results given here rely on the accuracy of the neutron interaction modeling in {\tt FLUKA}, and direct neutron test-beam measurements are highly desirable to validate the simulations. In this case, we note two important, complementary experimental efforts. The first is with the mini-CAPTAIN detector. This experiment already ran and collected data at LANL~\cite{Bhandari:2019rat}, in a neutron beam with energies between 100 and 800 MeV, but so far has only presented total cross section results. We encourage the collaboration to specifically analyze the distribution of the leading knockout proton energies.

The second is the calibration study that was conducted at ProtoDUNE this summer, using a 2.5 MeV pulsed neutron source. These measurements should help validate the neutron transport and capture model and we are eagerly awaiting the release of the results.

To this end, we simulate energy reconstruction expected from a neutron test beam. In Fig.~\ref{fig:neutron4panels}, we present results for neutrons of  initial energies of 0.1, 0.5, 1.0, and 2.0 GeV. One of the most striking observations here is that at high energies, $\gtrsim 1$ GeV, the histograms of reconstructed energy begin to look quite similar to those of the charged hadrons, shown earlier. We will return to this important point in Sec.~\ref{sect:discussion}.


\section{Effects of limited volume}
\label{sect:limitedvolume}

\begin{figure*}
	\begin{center}
        \includegraphics[width=0.32\textwidth]{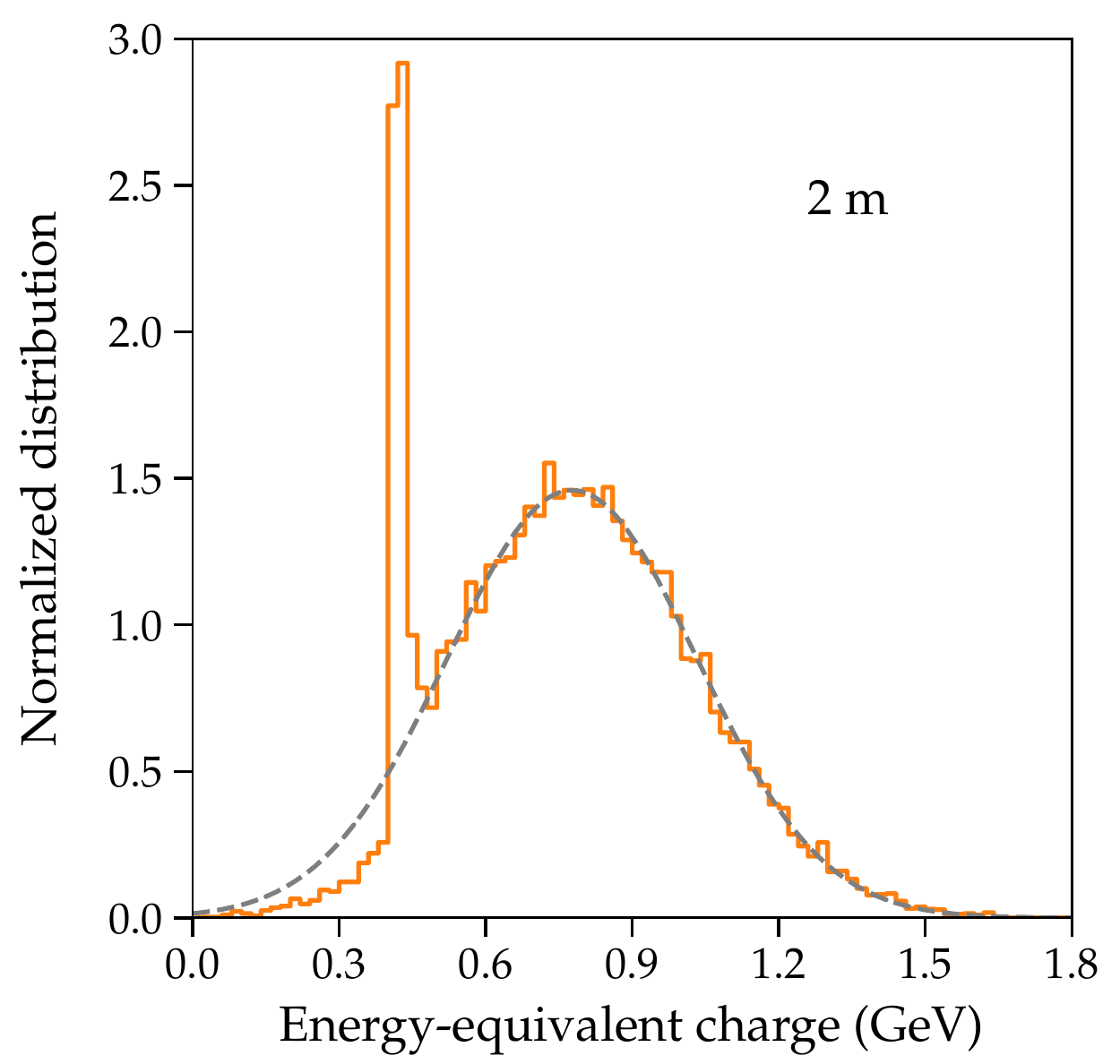}
        \includegraphics[width=0.32\textwidth]{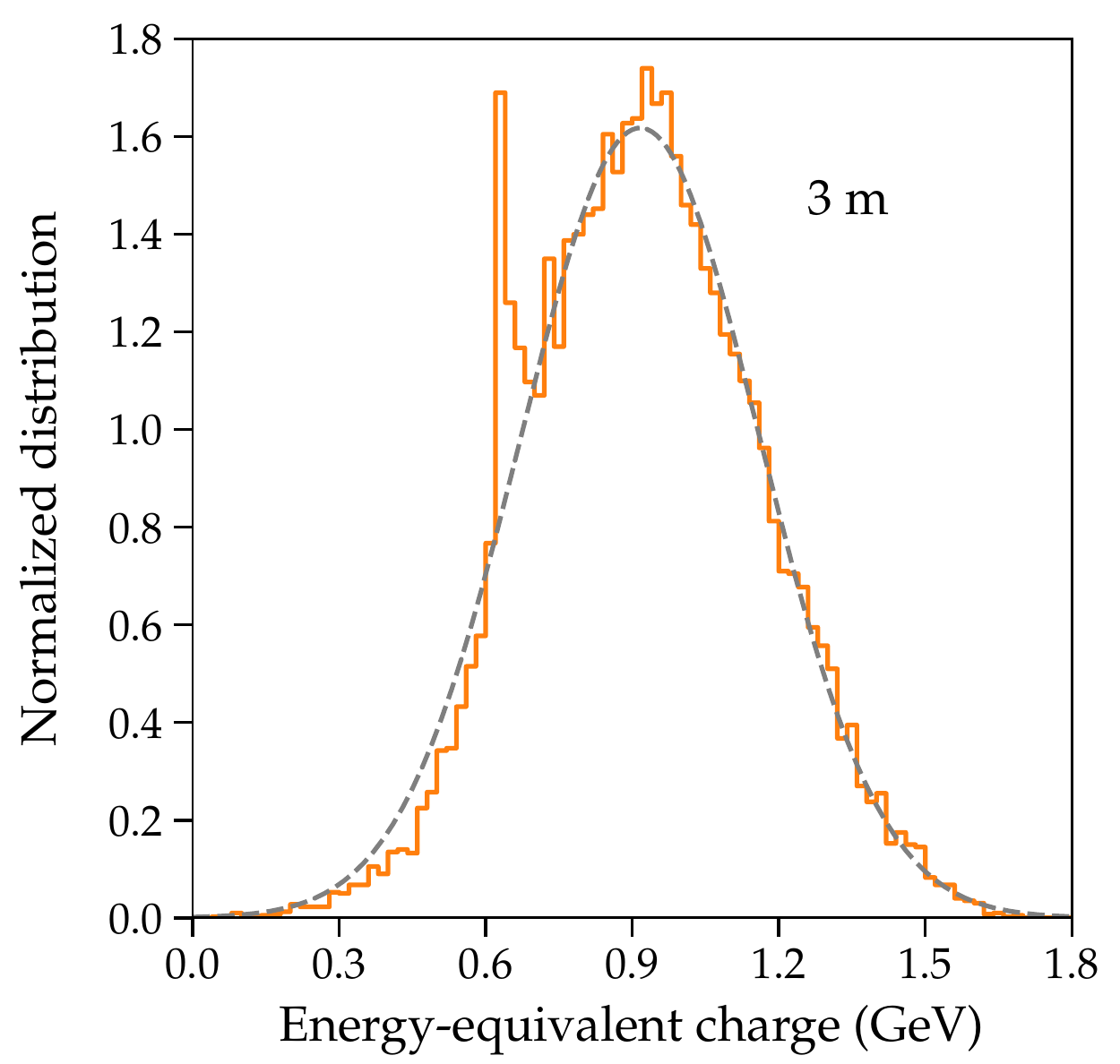}
        \includegraphics[width=0.32\textwidth]{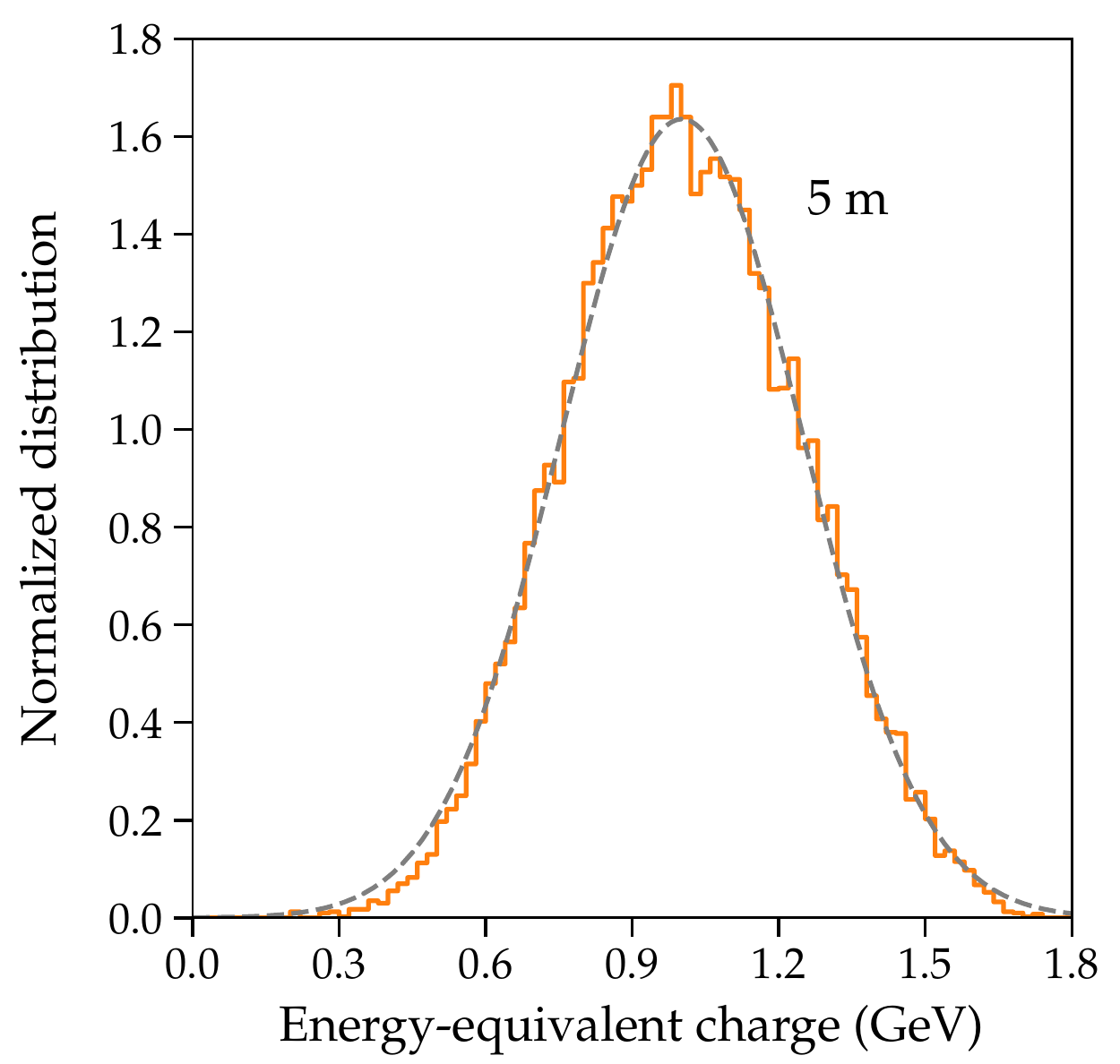}
        \includegraphics[width=0.32\textwidth]{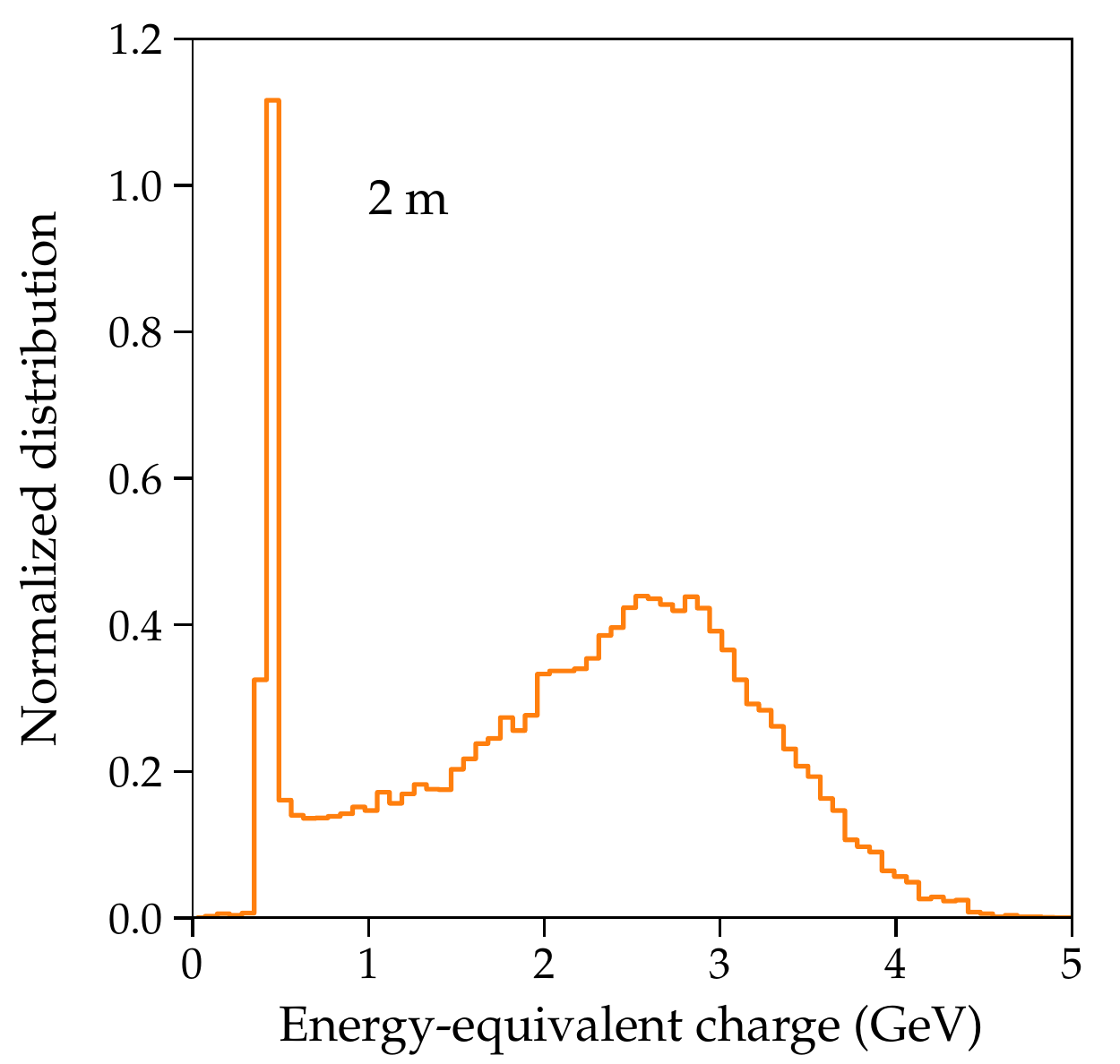}
        \includegraphics[width=0.32\textwidth]{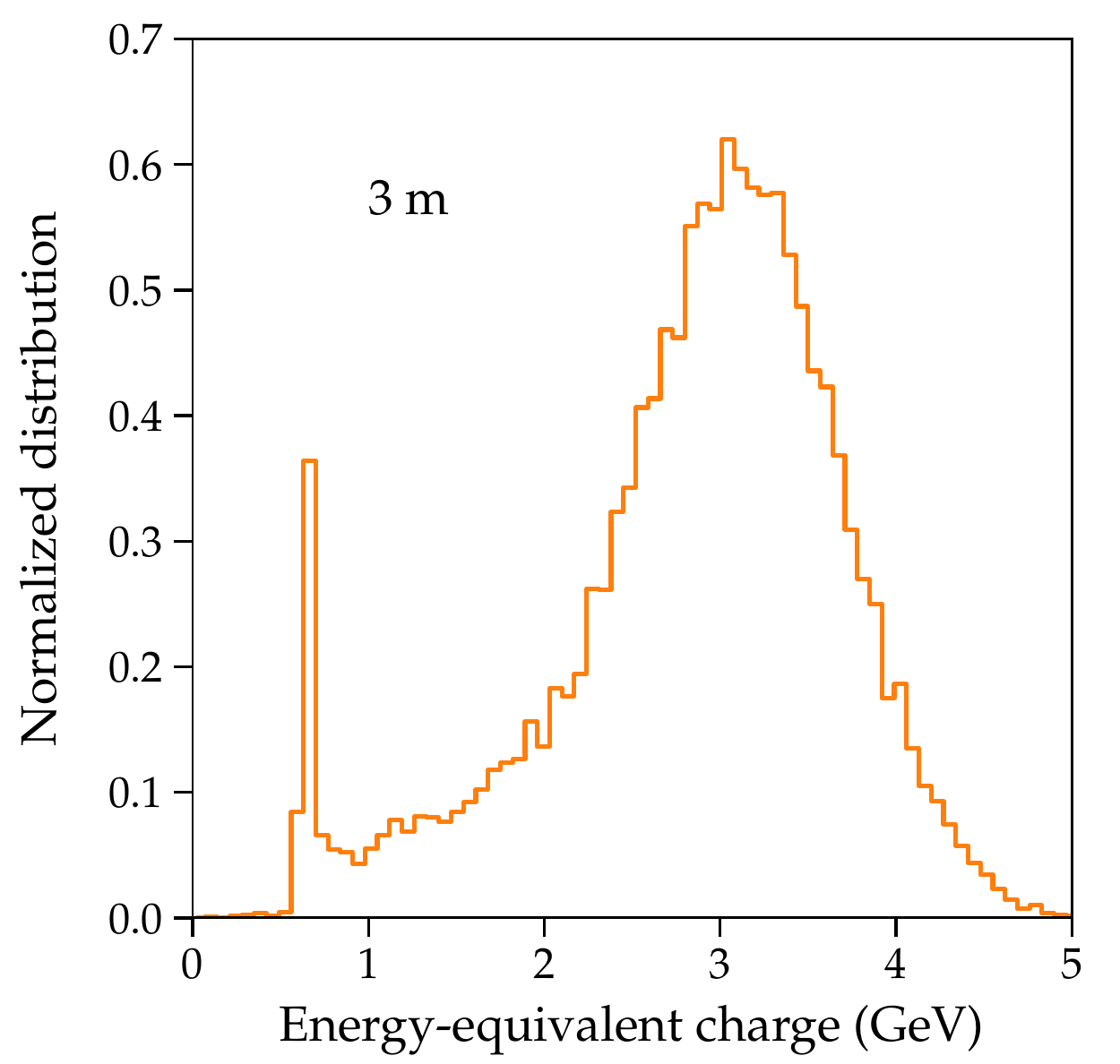}
        \includegraphics[width=0.32\textwidth]{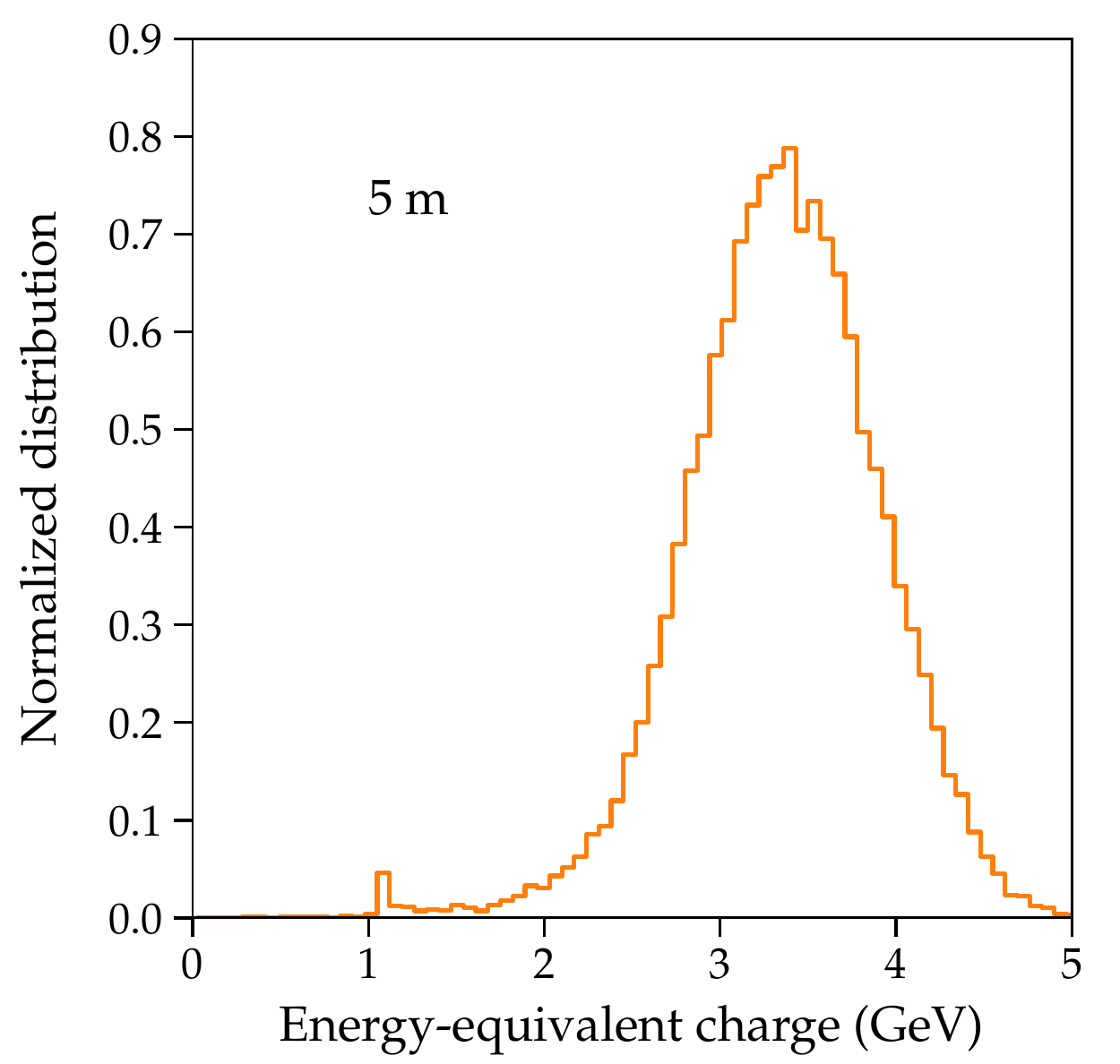}
        \caption{Distribution of ionization charges created by an injected proton in cubic volumes of length 2, 3, and 5~m. The top row corresponds to injected proton energy of 2 GeV; the bottom row, to proton energy of 7 GeV. The dashed curves in the top row show the corresponding Gaussian fits.}
        \label{fig:containment}
    \end{center}
\end{figure*}

As the next step, we will consider what happens if the detection volume is limited. This study has two motivations. From the practical side, such a situation could be realized in ProtoDUNE~\cite{Abi:2020mwi}, if one analyzes ionization charges collected in a single anode plane assembly, or light detected by a single light collection bar~\footnote{We thank Flavio Cavanna for bringing this possibility to our attention.}. It may also have implications for the design of near detectors, as we noted in Ref.~\cite{Friedland:2018vry}. From the conceptual point of view, we would like to understand how the spatial development of the events impacts the accuracy of calorimetric measurements. 

We consider proton beams with two initial energy values, 2 GeV and 7 GeV. The first case is motivated by the relevance to the DUNE experiment, where the neutrino energy varies in the $\sim$1--4 GeV range. The second one occurs in ProtoDUNE, where the test beam energies ran a range of values, including 7 GeV.

The simulation results are collected in Fig.~\ref{fig:containment}, where the volumes considered are 2 $\times$ 2 $\times$ 2 m, 3 $\times$ 3 $\times$ 3 m, and 5 $\times$ 5 $\times$ 5 m, left to right. The top row corresponds to injected protons of 2 GeV energy, and the bottom row shows the corresponding results for 7 GeV protons. In each case, we consider the method of total charge calorimetry and bin the simulation results in ``energy-equivalent charge", which is defined as energy lost by a minimally-ionizing muon that creates the same amount of ionization charge. Specifically, one ionization electron is counted as 23.6 eV of lost energy~\cite{Friedland:2018vry,Abi:2020mwi}.

We see that, while in the 5 $\times$ 5 $\times$ 5 m volume the charge distribution closely follows a simple Gaussian shape, the situation in the smaller volumes is more complicated. In addition to the scattered component, we also clearly see an unscattered one. Given the mean free path for hadronic interactions $\sim 80$ cm, the fraction of unscattered protons exiting the 2 $\times$ 2 $\times$ 2 m volume is $\sim\exp(-2/0.8)\sim8\%$, consistent with what is seen in the histogram.

The second relevant observation concerns the scattered component. The dashed curves in the top panels show the corresponding Gaussian fits. We see that, while the absolute width of the Gaussian stays approximately the same in all volumes, the center of the Gaussian moves to higher energies (charge) as the volume is increased. This indicates that in the smaller volumes the shower is not yet fully developed. For example, for 2 GeV injected protons, about 22\% of all ionization charges are created outside of the 2 $\times$ 2 $\times$ 2 m volume, and 8\% are created outside of the 3 $\times$ 3 $\times$ 3 m volume. 

\begin{figure}
	\begin{center}
        \includegraphics[width=\columnwidth]{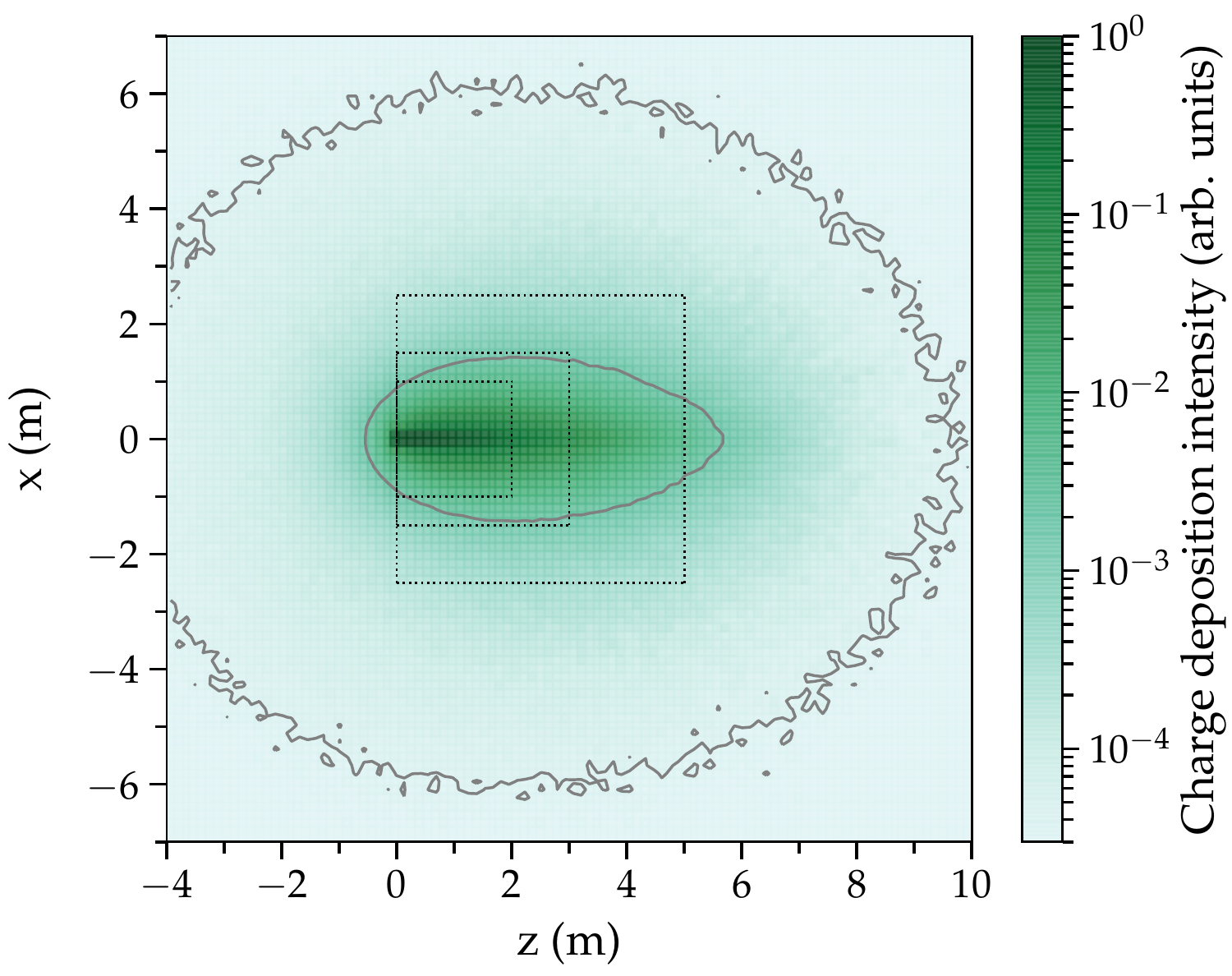}
        \caption{Distribution of ionization charges created by injecting $4\times10^5$ protons of 2 GeV kinetic energy at position $(0,0,0)$. The initial proton momenta point in the positive $z$ direction. All charges have been projected along the $y$ direction. The solid contours show the regions enclosing 95\% and 99\% of the total charge. The dashed lines show the 2 $\times$ 2 $\times$ 2 m, 3 $\times$ 3 $\times$ 3 m, and 5 $\times$ 5 $\times$ 5 m cubic volumes considered in Fig.~\ref{fig:containment}.}
        \label{fig:chargedist}
    \end{center}
\end{figure}

This is directly confirmed by examining the spatial distribution of the ionization charge in our simulation. Figure~\ref{fig:chargedist} shows the distribution of charges found after injecting $4 \times 10^5$ protons at position $(0,0,0)$. The initial proton kinetic energy is 2 GeV, and the momentum points along the $z$ direction. The $y$ coordinate has been suppressed, so that the graphics shows the charge projection onto the $(x,z)$ plane. The contours show the regions enclosing 95\% and 99\% of the total charge. The cubic volumes considered above are shown with dashed lines.

The smaller volumes clearly fail to enclose the full charge distribution. Even the 5-m box misses a few percent of the ionization charge. This extended charge ``halo'' is created mostly by diffusing neutrons. Interestingly, some of the charge lies in the backward direction (at negative $z$). This charge cannot be captured at ProtoDUNE, but may be detected in the DUNE far detector.

For 7 GeV injected protons, the effects of the limited volume are even more pronounced, as indicated by an extended shoulder between the unscattered spike and the peak of the scattered distribution.

This shows that behind seemingly simple Gaussian resolution curves seen in Sec.~\ref{sect:results} lies a complicated dynamical picture of shower development. The resolution of a detector may thus be affected by its geometry and other relevant considerations, such as requirements to fiducialize the detection volume to eliminate cosmic-ray-induced and other contamination.


\section{Discussion}
\label{sect:discussion}

\begin{figure*}
	\begin{center}
        \includegraphics[width=0.32\textwidth]{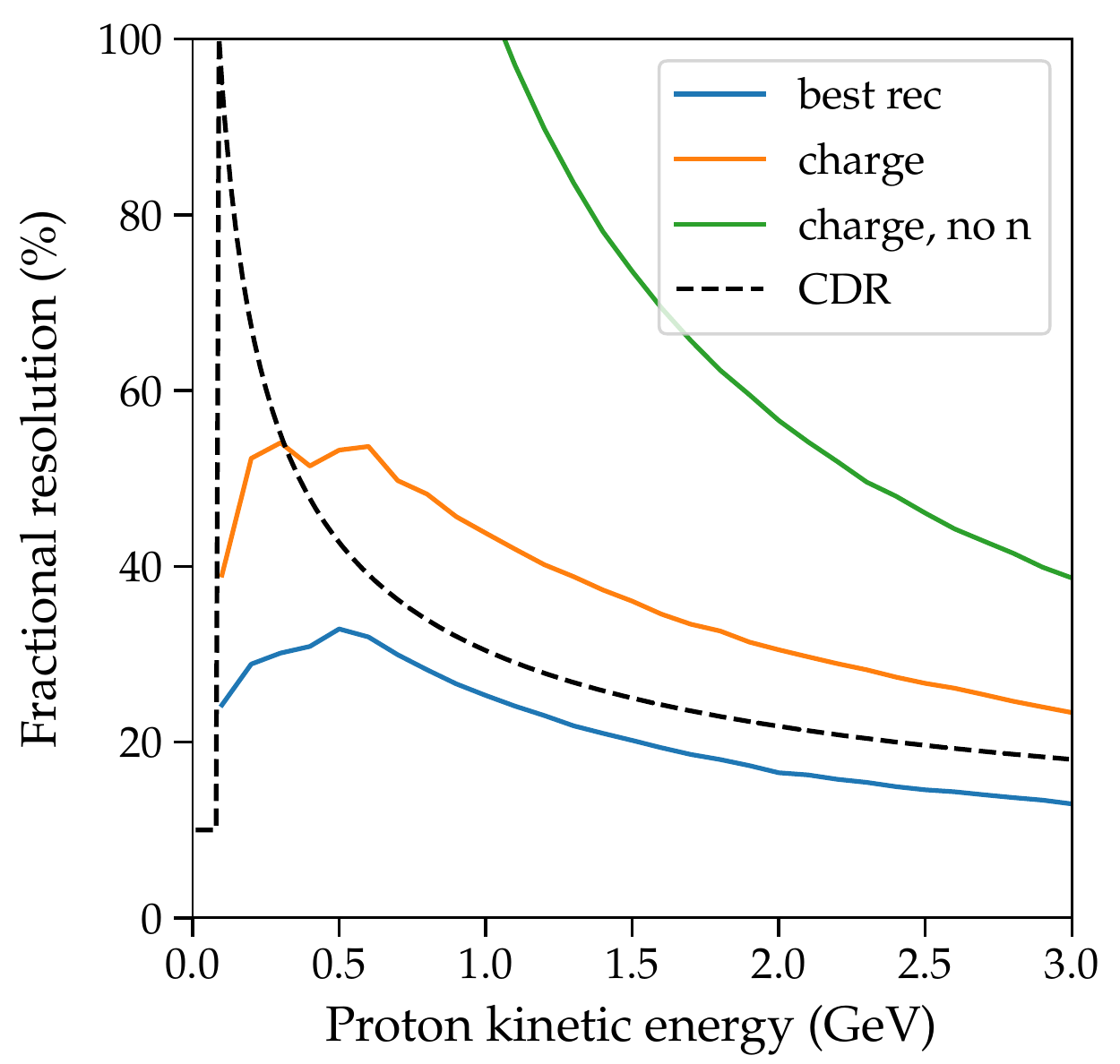}
        \includegraphics[width=0.32\textwidth]{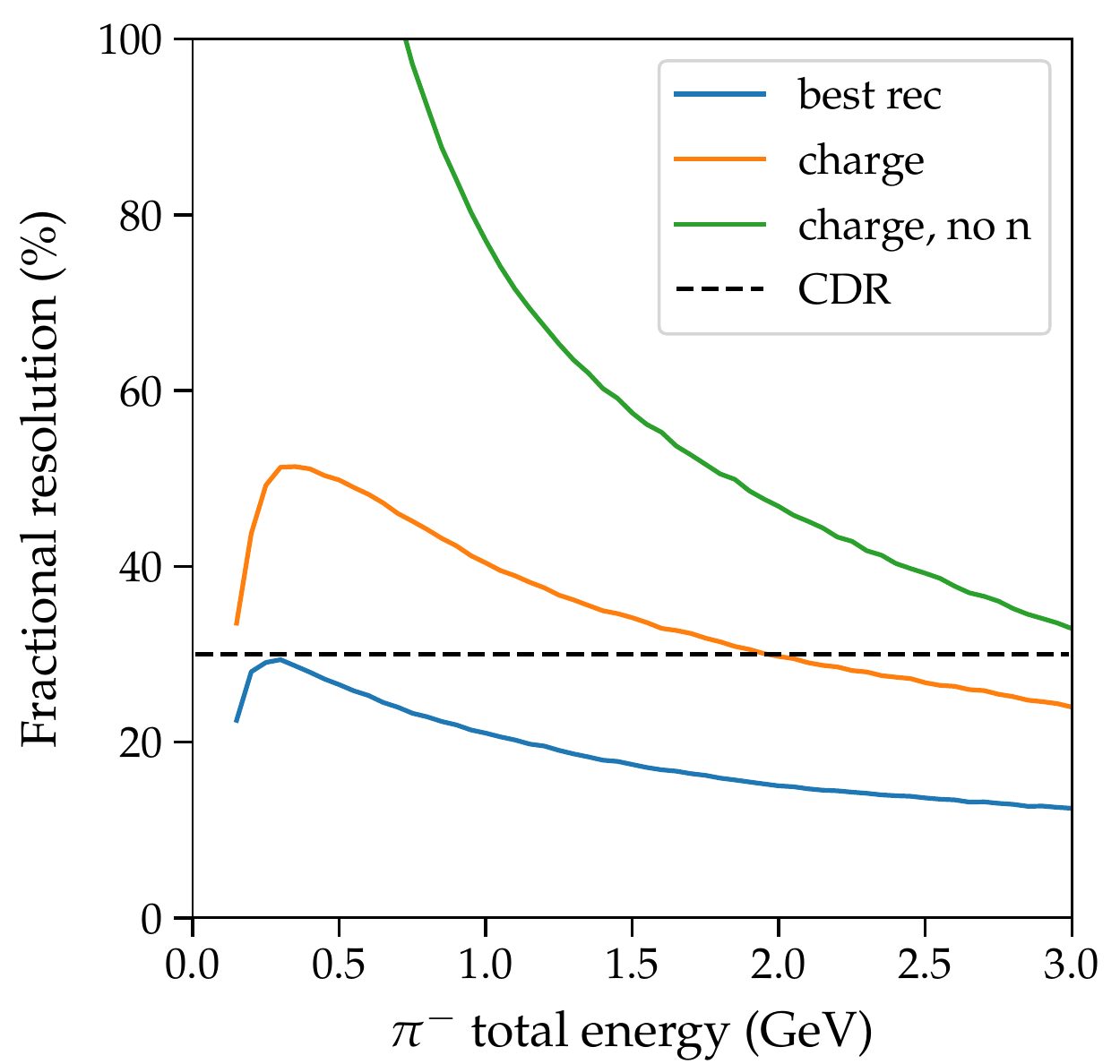}
        \includegraphics[width=0.32\textwidth]{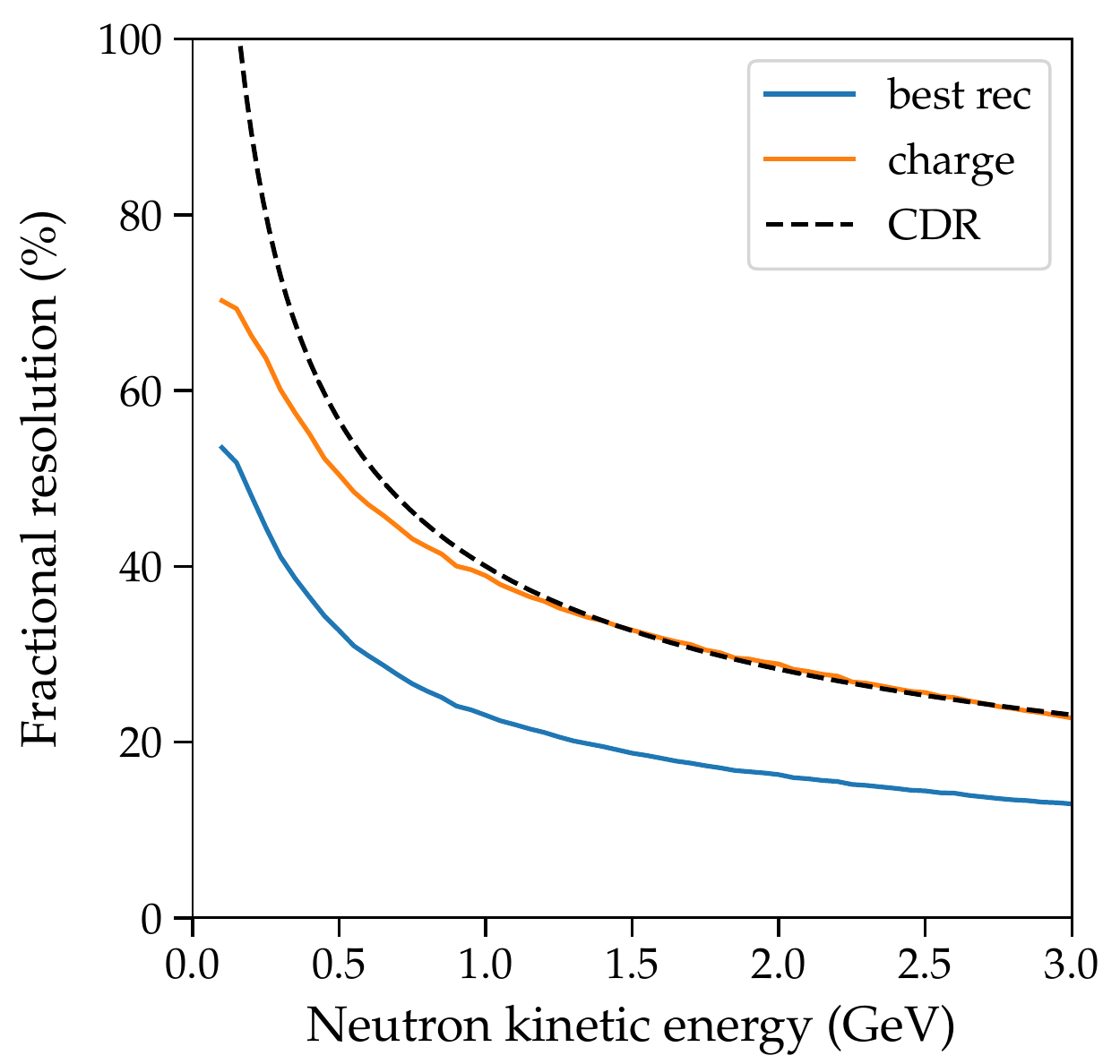}
        \caption{Simulated hadron energy resolution as a function of its true energy. Left to right: protons, negative pions, and neutrons.}
        \label{fig:protonresolution}
    \end{center}
\end{figure*}
 
The results of our large-volume simulations can be summarized by plotting the energy resolution for each particle type, as a function of energy. This is shown in Fig.~\ref{fig:protonresolution}, where injection energies up to 3 GeV are considered. The colored curves correspond to the three reconstruction scenarios we consider, as labeled. The dashed curves indicate the resolution assumed in the CDR document~\cite{Acciarri:2015uup,Acciarri:2016ooe}. 

We immediately see that the role of neutrons is absolutely crucial for the accuracy of charge hadron energy reconstruction: the green curves, which correspond to discarding all neutrons, show the resolution that is significantly worse than the other two cases. This is in line with what we already discussed in Sec.~\ref{sect:results} for specific energy values. Even though the average fraction of energy that goes into secondary neutrons is quite stable, about 20\%, the event-by-event variation of this fraction is very large~\cite{Friedland:2018vry}.

Let us now turn to the other two reconstruction scenarios. Notably, at sufficiently high energies, the fractional energy resolution is well fit by a $E^{-1/2}$ scaling law. Specifically, for protons we obtain $42\%/\sqrt{E}$ for the charge-only method and $25\%/\sqrt{E}$ for the best-reconstruction method. For charged pions, we find $42\%/\sqrt{E}$ for the charge-only method and $21\%/\sqrt{E}$ for the best-reconstruction method. For neutrons, the corresponding relationships are  $40\%/\sqrt{E}$ for the charge-only method and $23\%/\sqrt{E}$ for the best-reconstruction method. The first observation, therefore, is that at high energies the energy resolution performance is remarkably similar for each particle type.

The second observation is that the $E^{-1/2}$ law breaks down at lower energies, and the fractional resolution actually improves as the energy is decreased to 0.1 GeV. Let us discuss the underlying reasons for this behavior.

At the most basic level, liquid argon detectors operate as calorimeters, in which ionization charge deposited by particles created as a result of neutrino interactions is used to infer the total energy. Conversion from charge to energy involves, however, a number of steps that each introduce uncertainty. The size of this uncertainty depends on the amount of additional information gained in the reconstruction process. Let us summarize the relevant factors:

a.	For a given final-state track, the first consideration is its PID. Conversion from charge deposited along a track to energy involves correctly accounting for charges lost to recombination. The recombination correction is higher for slow-moving protons than for pions and muons of the same kinetic energies, as illustrated in Fig.~\ref{fig:recombination}. One might wonder whether the discrimination between neutral and charged pions, {\it i.e.}, between electromagnetic showers and tracks, also plays a role here. In fact, we explicitly checked that, in our simulation, the main impact of PID on the accuracy of energy reconstruction is through proton/charged pion discrimination.

b.	The next fundamental ingredient in the energy reconstruction of charged hadrons is their interactions in the medium. Indeed, once the particle type is identified, $dQ/dx$ along its trajectory can be reasonably well related to $dE/dx$, until the particle undergoes a hadronic interaction with a background argon nucleus. 

In hadronic collisions, the energy flow is affected by several processes:

i. Some energy is lost to the breakup of the target nucleus. Some can be emitted by de-excitation gamma rays, which create small charge deposits that may be detected with a varying degree of efficiency, depending on the detection thresholds.

ii.	Energy can be imparted to one or more hadrons, such as secondary pions created in the collision, nucleons knocked out of the nucleus, or a combination of pions and nucleons. For each secondary track, the accuracy of conversion from charge to energy loss again depends on whether PID information is available.

iii. Some of the knocked-out nucleons in the last step could be neutrons, and these present a special challenge, as discussed in Sec.~\ref{sect:neutrons}. They do not leave tracks and can dissipate energy by exciting and breaking up numerous argon nuclei, resulting in a spray of small charge deposits. They may also produce tertiary charged hadrons, which are likewise detached from the main event. Energy reconstruction depends on whether and how often such detached charge deposits can be identified with the main event.

Above all, the main conclusion here is this: the nature of the energy resolution is dictated by the frequency of hadronic collisions. Hadrons above 1 GeV (and their products) are expected to undergo multiple collisions. In this regime, the distribution of energy among the several channels becomes stochastic, and the reconstructed energy distribution approaches a Gaussian form. Notice that the widths of the Gaussians, which have been derived earlier, are found to be very similar for the three hadron types. They are controlled by the similar hadronic interaction rates. 

On the other hand, at lower energies, the interactions are only sporadic, and the distributions of reconstructed energies become more and more asymmetric. The Gaussian width prescriptions obtained at higher energies break down at these energies. For protons of $\sim$100~MeV energy, the high-energy Gaussian width fails dramatically. Instead, the energy can be reconstructed with very good accuracy, assuming good PID.

As a corollary, for protons and charged pions, the worst relative resolution occurs at energies of several hundred MeV, as seen in the graphs. We see that this behavior is not captured by the assumptions of the CDR (shown with dashed curves).

Given the crucial role of the hadronic interactions, it is essential that our predictions for them (made with {\tt FLUKA}) be directed tested with ProtoDUNE. This applies not only to the frequency of collisions, but also to the statistics of the final states produced. 

Let us now consider some important applications for our results. Consider two types of problems:
\begin{itemize}
    \item estimating the impact of various detector changes---such as gradually improving neutron detection efficiency, or improving PID; and,
    \item understanding the impact of various cross section uncertainties, especially the impact of several continuously varied parameters in the model.
\end{itemize}
For example, suppose one considers changes to the pion production model for neutrino-nucleon interactions, to reduce the tensions with the electron scattering data~\cite{Ankowski:2020qbe}. This adjustment may result in the modification of the properties of the hadronic final states~\cite{Ankowski:2019mfd}. One would like to be able to gauge the impact of these changes on neutrino energy reconstruction, without having to regenerate the full event simulation set after each incremental adjustment, which carries prohibitive computing costs. 

This calls for the need to build simplified codes, as we mentioned in the Introduction. Such codes would, instead of simulating full events in the detector, apply certain ``smearing'' prescriptions to the final-state particles output by the neutrino-nucleus event generator, in the spirit of {\tt FastMC}~\cite{Acciarri:2015uup,Acciarri:2016ooe}. Such a framework would give approximate answers to the questions of energy resolution and energy scale calibration, in response to various assumptions about cross section physics or detector performance. It can also be used to explore sensitivity to various new physics scenarios.

Our virtual test-beam simulations provide crucial input into such a framework. As we saw, it not only gives the width of the distribution of reconstructed energy but also specifies when the energy of a particle can be Gaussian smeared and when a different functional form must be used. 


\section{Conclusions}
\label{sec:conclusions}

In summary, the two main lessons of our investigations in this paper are as follows. First, the energy resolution of liquid argon time-projection chamber detectors strongly depends on the detector parameters and performance. Among the relevant factors are the detector geometry, which may impact event containment, and the quality of event reconstruction. In particular, the inability to reconstruct detached charge deposits due to neutrons leads to a large resolution penalty. 

Second, for hadrons with energies in the GeV range, the resulting distributions of reconstructed energies are often non-Gaussian. Namely:
\begin{itemize}
\item  With neutrons dropped, we consistently find a very non-Gaussian charge distribution, even when the detection volume is large.
\item  Conversely, in a limited volume (2 $\times$ 2 $\times$ 2 m), for high initial energy, we again get a non-Gaussian charge distribution, even with neutrons included.
\item  We have considered a total charge measurement with no PID corrections. In a large volume, with detached charges created by neutrons, the distribution starts approaching Gaussian at 2 GeV. 
\item  The best-case scenario is when the charges are collected over a large volume, neutron-induced charges are included and full PID corrections are implemented. In this case, the distribution is Gaussian even at 1 GeV.
\item Even in the best-case scenario, however, at low hadron energies the distribution is always non-Gaussian; this happens for proton energies $\lesssim0.6$~GeV and charged pion energies $\lesssim0.4$ GeV. 
\end{itemize}

We noted before~\cite{Friedland:2018vry}, that loss of information about a neutrino-induced event always leads to worsening of energy resolution. We clearly see this here, at the level of individual hadrons. We also see that the same loss of information---either by failure to contain the full event or by missing some particles---often leads to non-Gaussianity of the reconstructed energy distribution.

We see that the situation is quite different from the case of highly energetic particles, where the corresponding hadronic shower can fully develop. In that case, numerous statistical fluctuations combine to make the calorimetrically determined energy fluctuate in an approximately Gaussian manner. In the range of energies relevant to DUNE, however, $\mathcal{O}(1\mbox{ GeV})$, the shower may not be developed, as we have seen here. 

Our findings have two major applications. First, they can be directly applied to the analysis of the test-beam ProtoDUNE data. The comparison should make it possible to validate the parameters of the simulation framework, as well as help guide the analysis of the experimental data. 

Second, they have implications for how the physics reach of liquid argon experiments is assessed. In situations where one is interested in general estimates of sensitivity to Beyond-the-Standard-Model scenarios, it may be acceptable to approximate the detector response with simple Gaussian errors. However, when accurate modeling is required---for example, in studying sensitivity to specific oscillation parameters---detailed, realistic models of the near and far detector are required for the results to be credible.

We hope that the present study will help with constructing such detailed models.


\begin{acknowledgments}
We are grateful to Flavio Cavanna and Yun-Tse Tsai for useful discussions. We owe special thanks to the {\tt FLUKA} development team for their support with the package. Both authors are supported by the U.S. Department of Energy under Contract No. DE-AC02-76SF00515.
\end{acknowledgments}


\bibliography{larreferences}

\begin{thebibliography}{17}%
\makeatletter
\providecommand \@ifxundefined [1]{%
 \@ifx{#1\undefined}
}%
\providecommand \@ifnum [1]{%
 \ifnum #1\expandafter \@firstoftwo
 \else \expandafter \@secondoftwo
 \fi
}%
\providecommand \@ifx [1]{%
 \ifx #1\expandafter \@firstoftwo
 \else \expandafter \@secondoftwo
 \fi
}%
\providecommand \natexlab [1]{#1}%
\providecommand \enquote  [1]{``#1''}%
\providecommand \bibnamefont  [1]{#1}%
\providecommand \bibfnamefont [1]{#1}%
\providecommand \citenamefont [1]{#1}%
\providecommand \href@noop [0]{\@secondoftwo}%
\providecommand \href [0]{\begingroup \@sanitize@url \@href}%
\providecommand \@href[1]{\@@startlink{#1}\@@href}%
\providecommand \@@href[1]{\endgroup#1\@@endlink}%
\providecommand \@sanitize@url [0]{\catcode `\\12\catcode `\$12\catcode
  `\&12\catcode `\#12\catcode `\^12\catcode `\_12\catcode `\%12\relax}%
\providecommand \@@startlink[1]{}%
\providecommand \@@endlink[0]{}%
\providecommand \url  [0]{\begingroup\@sanitize@url \@url }%
\providecommand \@url [1]{\endgroup\@href {#1}{\urlprefix }}%
\providecommand \urlprefix  [0]{URL }%
\providecommand \Eprint [0]{\href }%
\providecommand \doibase [0]{http://dx.doi.org/}%
\providecommand \selectlanguage [0]{\@gobble}%
\providecommand \bibinfo  [0]{\@secondoftwo}%
\providecommand \bibfield  [0]{\@secondoftwo}%
\providecommand \translation [1]{[#1]}%
\providecommand \BibitemOpen [0]{}%
\providecommand \bibitemStop [0]{}%
\providecommand \bibitemNoStop [0]{.\EOS\space}%
\providecommand \EOS [0]{\spacefactor3000\relax}%
\providecommand \BibitemShut  [1]{\csname bibitem#1\endcsname}%
\let\auto@bib@innerbib\@empty
\bibitem [{\citenamefont {Abi}\ \emph {et~al.}(2020{\natexlab{a}})\citenamefont
  {Abi} \emph {et~al.}}]{Abi:2020evt}%
  \BibitemOpen
  \bibfield  {author} {\bibinfo {author} {\bibfnamefont {B.}~\bibnamefont
  {Abi}} \emph {et~al.} (\bibinfo {collaboration} {DUNE}),\ }\href@noop {} {\
  (\bibinfo {year} {2020}{\natexlab{a}})},\ \Eprint
  {http://arxiv.org/abs/2002.03005} {arXiv:2002.03005 [hep-ex]} \BibitemShut
  {NoStop}%
\bibitem [{\citenamefont {Friedland}\ and\ \citenamefont
  {Li}(2019)}]{Friedland:2018vry}%
  \BibitemOpen
  \bibfield  {author} {\bibinfo {author} {\bibfnamefont {A.}~\bibnamefont
  {Friedland}}\ and\ \bibinfo {author} {\bibfnamefont {S.~W.}\ \bibnamefont
  {Li}},\ }\href {\doibase 10.1103/PhysRevD.99.036009} {\bibfield  {journal}
  {\bibinfo  {journal} {Phys. Rev. D}\ }\textbf {\bibinfo {volume} {99}},\
  \bibinfo {pages} {036009} (\bibinfo {year} {2019})},\ \Eprint
  {http://arxiv.org/abs/1811.06159} {arXiv:1811.06159 [hep-ph]} \BibitemShut
  {NoStop}%
\bibitem [{\citenamefont {Abi}\ \emph {et~al.}(2020{\natexlab{b}})\citenamefont
  {Abi} \emph {et~al.}}]{Abi:2020mwi}%
  \BibitemOpen
  \bibfield  {author} {\bibinfo {author} {\bibfnamefont {B.}~\bibnamefont
  {Abi}} \emph {et~al.} (\bibinfo {collaboration} {DUNE}),\ }\href@noop {} {\
  (\bibinfo {year} {2020}{\natexlab{b}})},\ \Eprint
  {http://arxiv.org/abs/2007.06722} {arXiv:2007.06722 [physics.ins-det]}
  \BibitemShut {NoStop}%
\bibitem [{\citenamefont {Acciarri}\ \emph {et~al.}(2015)\citenamefont
  {Acciarri} \emph {et~al.}}]{Acciarri:2015uup}%
  \BibitemOpen
  \bibfield  {author} {\bibinfo {author} {\bibfnamefont {R.}~\bibnamefont
  {Acciarri}} \emph {et~al.} (\bibinfo {collaboration} {DUNE}),\ }\href@noop {}
  {\  (\bibinfo {year} {2015})},\ \Eprint {http://arxiv.org/abs/1512.06148}
  {arXiv:1512.06148 [physics.ins-det]} \BibitemShut {NoStop}%
\bibitem [{\citenamefont {Acciarri}\ \emph {et~al.}(2016)\citenamefont
  {Acciarri} \emph {et~al.}}]{Acciarri:2016ooe}%
  \BibitemOpen
  \bibfield  {author} {\bibinfo {author} {\bibfnamefont {R.}~\bibnamefont
  {Acciarri}} \emph {et~al.} (\bibinfo {collaboration} {DUNE}),\ }\href@noop {}
  {\  (\bibinfo {year} {2016})},\ \Eprint {http://arxiv.org/abs/1601.02984}
  {arXiv:1601.02984 [physics.ins-det]} \BibitemShut {NoStop}%
\bibitem [{\citenamefont {B{\"o}hlen\textbf{}}\ \emph
  {et~al.}(2014)\citenamefont {B{\"o}hlen\textbf{}}, \citenamefont {Cerutti},
  \citenamefont {Chin}, \citenamefont {Fass{\`o}}, \citenamefont {Ferrari},
  \citenamefont {Ortega}, \citenamefont {Mairani}, \citenamefont {Sala},
  \citenamefont {Smirnov},\ and\ \citenamefont {Vlachoudis}}]{Bohlen:2014buj}%
  \BibitemOpen
  \bibfield  {author} {\bibinfo {author} {\bibfnamefont {T.~T.}\ \bibnamefont
  {B{\"o}hlen\textbf{}}}, \bibinfo {author} {\bibfnamefont {F.}~\bibnamefont
  {Cerutti}}, \bibinfo {author} {\bibfnamefont {M.~P.~W.}\ \bibnamefont
  {Chin}}, \bibinfo {author} {\bibfnamefont {A.}~\bibnamefont {Fass{\`o}}},
  \bibinfo {author} {\bibfnamefont {A.}~\bibnamefont {Ferrari}}, \bibinfo
  {author} {\bibfnamefont {P.~G.}\ \bibnamefont {Ortega}}, \bibinfo {author}
  {\bibfnamefont {A.}~\bibnamefont {Mairani}}, \bibinfo {author} {\bibfnamefont
  {P.~R.}\ \bibnamefont {Sala}}, \bibinfo {author} {\bibfnamefont
  {G.}~\bibnamefont {Smirnov}}, \ and\ \bibinfo {author} {\bibfnamefont
  {V.}~\bibnamefont {Vlachoudis}},\ }\href {\doibase 10.1016/j.nds.2014.07.049}
  {\bibfield  {journal} {\bibinfo  {journal} {Nucl. Data Sheets}\ }\textbf
  {\bibinfo {volume} {120}},\ \bibinfo {pages} {211} (\bibinfo {year}
  {2014})}\BibitemShut {NoStop}%
\bibitem [{\citenamefont {Ferrari}\ \emph {et~al.}()\citenamefont {Ferrari},
  \citenamefont {Sala}, \citenamefont {Fass{\`o}},\ and\ \citenamefont
  {Ranft}}]{Ferrari:2005zk}%
  \BibitemOpen
  \bibfield  {author} {\bibinfo {author} {\bibfnamefont {A.}~\bibnamefont
  {Ferrari}}, \bibinfo {author} {\bibfnamefont {P.~R.}\ \bibnamefont {Sala}},
  \bibinfo {author} {\bibfnamefont {A.}~\bibnamefont {Fass{\`o}}}, \ and\
  \bibinfo {author} {\bibfnamefont {J.}~\bibnamefont {Ranft}},\ }\href@noop {}
  {\ }\bibinfo {note} {{FLUKA: A multi-particle transport code; CERN-2005-010;
  SLAC-R-773; INFN-TC-05-11}}\BibitemShut {NoStop}%
\bibitem [{\citenamefont {Acciarri}\ \emph {et~al.}(2019)\citenamefont
  {Acciarri} \emph {et~al.}}]{Acciarri:2018myr}%
  \BibitemOpen
  \bibfield  {author} {\bibinfo {author} {\bibfnamefont {R.}~\bibnamefont
  {Acciarri}} \emph {et~al.} (\bibinfo {collaboration} {ArgoNeuT}),\ }\href
  {\doibase 10.1103/PhysRevD.99.012002} {\bibfield  {journal} {\bibinfo
  {journal} {Phys. Rev. D}\ }\textbf {\bibinfo {volume} {99}},\ \bibinfo
  {pages} {012002} (\bibinfo {year} {2019})},\ \Eprint
  {http://arxiv.org/abs/1810.06502} {arXiv:1810.06502 [hep-ex]} \BibitemShut
  {NoStop}%
\bibitem [{\citenamefont {Sorel}(2014)}]{Sorel:2014rka}%
  \BibitemOpen
  \bibfield  {author} {\bibinfo {author} {\bibfnamefont {M.}~\bibnamefont
  {Sorel}},\ }\href {\doibase 10.1088/1748-0221/9/10/P10002} {\bibfield
  {journal} {\bibinfo  {journal} {JINST}\ }\textbf {\bibinfo {volume} {9}},\
  \bibinfo {pages} {P10002} (\bibinfo {year} {2014})},\ \Eprint
  {http://arxiv.org/abs/1405.0848} {arXiv:1405.0848 [physics.ins-det]}
  \BibitemShut {NoStop}%
\bibitem [{\citenamefont {De~Romeri}\ \emph {et~al.}(2016)\citenamefont
  {De~Romeri}, \citenamefont {Fernandez-Martinez},\ and\ \citenamefont
  {Sorel}}]{DeRomeri:2016qwo}%
  \BibitemOpen
  \bibfield  {author} {\bibinfo {author} {\bibfnamefont {V.}~\bibnamefont
  {De~Romeri}}, \bibinfo {author} {\bibfnamefont {E.}~\bibnamefont
  {Fernandez-Martinez}}, \ and\ \bibinfo {author} {\bibfnamefont
  {M.}~\bibnamefont {Sorel}},\ }\href {\doibase 10.1007/JHEP09(2016)030}
  {\bibfield  {journal} {\bibinfo  {journal} {JHEP}\ }\textbf {\bibinfo
  {volume} {09}},\ \bibinfo {pages} {030} (\bibinfo {year} {2016})},\ \Eprint
  {http://arxiv.org/abs/1607.00293} {arXiv:1607.00293 [hep-ph]} \BibitemShut
  {NoStop}%
\bibitem [{\citenamefont {Grant}\ and\ \citenamefont
  {Yang}(2017)}]{Grant2018DPFtalk}%
  \BibitemOpen
  \bibfield  {author} {\bibinfo {author} {\bibfnamefont {N.}~\bibnamefont
  {Grant}}\ and\ \bibinfo {author} {\bibfnamefont {T.}~\bibnamefont {Yang}},\
  }in\ \href@noop {} {\emph {\bibinfo {booktitle} {talk at the DPF meeting}}}\
  (\bibinfo {year} {March 8, 2017})\ \bibinfo {note}
  {\url{https://indico.fnal.gov/event/11999/contribution/275}}\BibitemShut
  {NoStop}%
\bibitem [{\citenamefont {Castiglioni}\ \emph {et~al.}(2020)\citenamefont
  {Castiglioni}, \citenamefont {Foreman}, \citenamefont {Lepetic},
  \citenamefont {Littlejohn}, \citenamefont {Malaker},\ and\ \citenamefont
  {Mastbaum}}]{Castiglioni:2020tsu}%
  \BibitemOpen
  \bibfield  {author} {\bibinfo {author} {\bibfnamefont {W.}~\bibnamefont
  {Castiglioni}}, \bibinfo {author} {\bibfnamefont {W.}~\bibnamefont
  {Foreman}}, \bibinfo {author} {\bibfnamefont {I.}~\bibnamefont {Lepetic}},
  \bibinfo {author} {\bibfnamefont {B.}~\bibnamefont {Littlejohn}}, \bibinfo
  {author} {\bibfnamefont {M.}~\bibnamefont {Malaker}}, \ and\ \bibinfo
  {author} {\bibfnamefont {A.}~\bibnamefont {Mastbaum}},\ }\href@noop {} {\
  (\bibinfo {year} {2020})},\ \Eprint {http://arxiv.org/abs/2006.14675}
  {arXiv:2006.14675 [physics.ins-det]} \BibitemShut {NoStop}%
\bibitem [{\citenamefont {{McroBooNE
  Collaboration}}(2018)}]{MicroBooNE:2018jag}%
  \BibitemOpen
  \bibfield  {author} {\bibinfo {author} {\bibnamefont {{McroBooNE
  Collaboration}}},\ }\href {\doibase 10.2172/1573057} {\enquote {\bibinfo
  {title} {{Study of Reconstructed 39Ar Beta Decays at the MicroBooNE
  Detector}},}\ }\bibinfo {howpublished} {MICROBOONE-NOTE-1050-PUB,
  doi:10.2172/1573057} (\bibinfo {year} {2018})\BibitemShut {NoStop}%
\bibitem [{\citenamefont {Bhandari}\ \emph {et~al.}(2019)\citenamefont
  {Bhandari} \emph {et~al.}}]{Bhandari:2019rat}%
  \BibitemOpen
  \bibfield  {author} {\bibinfo {author} {\bibfnamefont {B.}~\bibnamefont
  {Bhandari}} \emph {et~al.} (\bibinfo {collaboration} {CAPTAIN}),\ }\href
  {\doibase 10.1103/PhysRevLett.123.042502} {\bibfield  {journal} {\bibinfo
  {journal} {Phys. Rev. Lett.}\ }\textbf {\bibinfo {volume} {123}},\ \bibinfo
  {pages} {042502} (\bibinfo {year} {2019})},\ \Eprint
  {http://arxiv.org/abs/1903.05276} {arXiv:1903.05276 [hep-ex]} \BibitemShut
  {NoStop}%
\bibitem [{Note1()}]{Note1}%
  \BibitemOpen
  \bibinfo {note} {We thank Flavio Cavanna for bringing this possibility to our
  attention.}\BibitemShut {Stop}%
\bibitem [{\citenamefont {Ankowski}\ and\ \citenamefont
  {Friedland}(2020)}]{Ankowski:2020qbe}%
  \BibitemOpen
  \bibfield  {author} {\bibinfo {author} {\bibfnamefont {A.~M.}\ \bibnamefont
  {Ankowski}}\ and\ \bibinfo {author} {\bibfnamefont {A.}~\bibnamefont
  {Friedland}},\ }\href {\doibase 10.1103/PhysRevD.102.053001} {\bibfield
  {journal} {\bibinfo  {journal} {Phys. Rev. D}\ }\textbf {\bibinfo {volume}
  {102}},\ \bibinfo {pages} {053001} (\bibinfo {year} {2020})},\ \Eprint
  {http://arxiv.org/abs/2006.11944} {arXiv:2006.11944 [hep-ph]} \BibitemShut
  {NoStop}%
\bibitem [{\citenamefont {Ankowski}\ \emph {et~al.}(2020)\citenamefont
  {Ankowski}, \citenamefont {Friedland}, \citenamefont {Li}, \citenamefont
  {Moreno}, \citenamefont {Schuster}, \citenamefont {Toro},\ and\ \citenamefont
  {Tran}}]{Ankowski:2019mfd}%
  \BibitemOpen
  \bibfield  {author} {\bibinfo {author} {\bibfnamefont {A.~M.}\ \bibnamefont
  {Ankowski}}, \bibinfo {author} {\bibfnamefont {A.}~\bibnamefont {Friedland}},
  \bibinfo {author} {\bibfnamefont {S.~W.}\ \bibnamefont {Li}}, \bibinfo
  {author} {\bibfnamefont {O.}~\bibnamefont {Moreno}}, \bibinfo {author}
  {\bibfnamefont {P.}~\bibnamefont {Schuster}}, \bibinfo {author}
  {\bibfnamefont {N.}~\bibnamefont {Toro}}, \ and\ \bibinfo {author}
  {\bibfnamefont {N.}~\bibnamefont {Tran}},\ }\href {\doibase
  10.1103/PhysRevD.101.053004} {\bibfield  {journal} {\bibinfo  {journal}
  {Phys. Rev. D}\ }\textbf {\bibinfo {volume} {101}},\ \bibinfo {pages}
  {053004} (\bibinfo {year} {2020})},\ \Eprint
  {http://arxiv.org/abs/1912.06140} {arXiv:1912.06140 [hep-ph]} \BibitemShut
  {NoStop}%
\end{thebibliography}%

\end{document}